\documentclass[preprint,nofootinbib]{revtex4-1}

\usepackage{graphicx}

\usepackage{amssymb}
\usepackage{amsmath}

\begin{document}
\sloppy

\title{Transfer Operators and Topological Field Theory
}


\author{Igor V. Ovchinnikov}



\affiliation{Electrical Engineering Department, University of California at Los Angeles, Los Angeles, CA 90095 USA}
\email{igor.vlad.ovchinnikov@gmail.com}



\begin{abstract}
The transfer operator (TO) formalism of the dynamical systems (DS) theory is reformulated here in terms of the recently proposed supersymetric theory of stochastic differential equations (SDE). It turns out that the stochastically generalized TO (GTO) of the DS theory is the finite-time Fokker-Planck evolution operator. As a result comes the supersymmetric trivialization of the so-called sharp trace and sharp determinant of the GTO, with the former being the Witten index, which is also the stochastic generalization of the Lefschetz index so that it equals the Euler characteristic of the (closed) phase space for any flow vector field, noise metric, and temperature. The enabled possibility to apply the spectral theorems of the DS theory to the Fokker-Planck operators allows to extend the previous picture of the spontaneous topological supersymmetry (Q-symmetry) breaking onto the situations with negative ground state's attenuation rate. The later signifies the exponential growth of the number of periodic solutions/orbits in the large time limit, which is the unique feature of chaotic behavior proving that the spontaneous breakdown of Q-symmetry is indeed the field-theoretic definition and stochastic generalization of the concept of deterministic chaos. In addition, the previously proposed low-temperature classification of SDEs, i.e., thermodynamic equilibrium / noise-induced chaos ((anti)instanton condensation, intermittent) / ordinary chaos (non-integrability of the flow vector field), is complemented by the discussion of the high-temperature regime where the sharp boundary between the noise-induced and ordinary chaotic phases must smear out into a crossover, and at even higher temperatures the Q-symmetry is restored. The Weyl quantization is discussed in the context of the Ito-Stratonovich dilemma.
\end{abstract}

\maketitle

\section{Introduction}
\label{Intro}
Soon after the proposition of the Parisi-Sourlas stochastic quantization procedure \cite{ParisiSourlas} for Langevin stochastic (partial) differential equations (SDEs), it was realized that being applied to a general form SDE, the Parisi-Sourlas technique leads to a model with at least one supersymmetry (see, \emph{e.g.}, \cite{Gaw}). This supersymmetry was later identified as a definitive feature of the Witten-type topological or cohomological field theories. \cite{Anselmi,Atiyah,TFT,Blau,Frenkel,Labastida,WittenTFT1,WittenTFT2} Since then, further progress in the studies of the connection between this topological supersymmetry and the general form SDEs was relatively slow (see, however, \cite{Kramers}) perhaps because in the general case (unlike in the Langevin case) the evolution operator is pseudo-Hermitian (it has imaginary eigenvalues), whereas the theory of the pseudo-Hermitian evolution operators (in the context of quantum theory) would not appear until around a decade ago. \cite{Mostafazadeh1,Mostafazadeh2}.

Yet another part of the story of this paper lays in the domain of the theory of deterministic chaos also known as non-integrability in the sense of dynamical systems (DS) theory \footnote{There are more than one versions of the concept of integrability in Mathematics. The integrability in the sense of DS theory is that the unstable manifolds of the flow vector field provide well-defined foliations of the phase space.} - one of the most fascinating known nonlinear dynamical phenomena. Its history is over a century old and can be traced back to the work by Poincar\'e on the three-body celestial dynamics (see, \emph{e.g.}, Ref. \cite{EarlyChaos}  and Refs. therein). At this, the classical DS theory has not offered yet an explanation/derivation of the most notorious chaotic property of the ultimate sensitivity to initial conditions also known as the "butterfly effect" (for the butterfly effect see, \emph{e.g.}, \cite{ChaosAtFifty} and Refs. therein). This is true even for deterministic DSs not to mention stochastic DSs.

Stochastic DSs, in turn, are of primary interest because all natural DSs are never completely isolated from their environments and thus always experience the influence from external stochastic noise. Even though previous approaches to stochastic dynamics provided many important insights, \cite{Arnold,KapitaniakBook,Hida,Baxendale} some of the fundamental questions remained unanswered. For example, there existed no rigorous stochastic generalization of the concept of deterministic chaos.

The theory that connects the above mentioned previous mathematical developments and provides answers to some of the open questions has appeared recently \cite{Mine2,Mine3} as a result of the conjecture \cite{Mine1} that chaotic (stochastic) dynamical behavior\footnote{Or rather one type of stochastic chaos known previously as self-organized criticality that corresponds in terminology of this paper to the noise-induced chaos, see Sec.\ref{GeneralPhaseDiagram}} may as well turn out to be the phenomenon of the spontaneous breakdown of topological sypersymmetry that all SDEs possess. 

This approximation-free and coordinate-free supersymmetric theory of SDEs or stochastics (STS), may find multiple applications because SDEs is probably the most fundamental class of models in all of science and engineering. Indeed, everything in Nature above the scale of quantum degeneracy/coherence is an SDE, whereas in quantum models\footnote{A Schr\"odinger equation is also a partial differential (Hamilton) equation that can be formally viewed as an infinite-dimensional deterministic (conservative) dynamical system. The STS approach to quantum mechanics is unlikely, however, to lead to any additional new knowledge about quantum dynamics.} SDEs can be used indirectly. \cite{Ringel} Among potential applications of the STS, its most interesting use is in the context of the DS theory and statistical physics as we discuss next.

The STS offers a rigorous definition of dynamical chaos that both explains the butterfly effect and works just as well for stochastic DSs: chaotic dynamics is the phenomenon of the spontaneous breakdown of topological supersymmetry. According to the Goldstone theorem, DSs with spontaneously broken topological supersymmetry must always exhibit "chaotic" long-range order/memory/correlations. The later reveals itself through such well-established concepts and natural phenomena as the butterfly effect, 1/f or pink/flicker noise, algebraic statistics of instantonic processes including earthquakes, solar flares, gamma-ray-bursts etc.

In fact, that the onset of chaos is a transition of some sort has been well known before. This understanding was actually the basis for the concept of "universality in chaos". \cite{Univ_Chao} In addition, it has also been known that the transition into chaos is of topological origin. At the transition, fractal attractors that are not well-defined topological manifolds show up. Furthermore, in three-dimensional phase spaces, fractal attractors consist of infinite number of unstable periodic orbits with arbitrary large periods and with nontrivial linking numbers. \cite{ChaosTopology} In the light of the above, the topological supersymmetry breaking picture is the long overdue field-theoretic explanation and stochastic generalization of the phenomenon of deterministic chaos.

Another fundamental physical concept that the STS provides with a rigorous definition is that of the thermodynamic equilibrium (TE). A stochastic DS is said to be at its TE if after long enough temporal evolution it can be described by a stationary total probability distribution (\emph{e.g.}, Gibbs distribution) and, in particular, loses all the memory of its initial conditions (no butterfly effect). As we will discuss in Sec.\ref{Discussion}, from the STS standpoint, DSs in TE are those with unbroken topological supersymmetry. In this case, (one of) the ground state(s) is the steady state (zero-eigenvalue) total probability distribution. As it turns out, such state always exist for physically meaningful models. Thus, the true essence of the TE in not in the existence of this TE state (this view can be often encountered in the literature) but rather in whether the TE state is among the ground states of the model or not.

With the provision of the definitions for the stochastic chaos and the TE, the STS offers a unified framework for physical statistics and dynamics: depending on whether the topological supersymmetry is spontaneously broken or not, all stochastic DSs are divided into chaotic ones and those at TE. At this, only DSs at TE admit the statistical/thermodynamic description. In order to capture the essence of chaotic behavior, one must construct a model-specific low-energy effective theory of gapless Fadeev-Popov goldstinos that represent "unthermalized" (unstable) variables \footnote{In deterministic dissipative continuous-time DSs, unstable variables correspond to positive Lyapuniov exponents.} of the ground state of a chaotic DS.

The STS may also be interesting from purely field-theoretic point of view. Firstly, through this theory, the ChTs may find multiple applications. Secondary, there are very few mechanisms that may lead to the spontaneous breakdown of supersymmetries, \cite{LectSusyBrk} which is actually the primary reason for the introduction of the concept of soft and/or explicit supersymmetry breaking (see Ref. \cite{softsusybreaking} and Refs. therein). The STS provides yet another mechanism of the spontaneous breakdown of a supersymmetry (topological supersymmetry in this case). Topological supersymmetry of chaotic deterministic DSs is spontaneously broken by  non-integrability of the flow vector field, \emph{i.e.}, by  fractal or strange attractors/invariant manifolds. Such mechanism does not have analogues in the high-energy physics models, in which the mean-field or tree-level vacua,  \emph{i.e.}, the analogues of the deterministic attractors/invariant manifolds of the DS theory, are well-defined (integer-dimensional) topological manifolds in the phase space. Interestingly enough, the chaotic ground state's eigenvalue of the Fokker-Planck operator is negative (it represents the rate of the exponential growth of the number of periodic solutions/orbits in chaotic DS, see Sec. \ref{SecChaos} below). This situation is in a sense opposite to the supersymmetry breaking picture in the high-energy physics models where the ground state eigenvalue must be positive in order for a supersymmetry to be spontaneously broken.

The Fokker-Planck evolution operators of the STS are non-Hermitian. Such evolution operators have been the subject of active scientific investigation recently. \cite{Bender1,Bender2,Mostafazadeh1,Mostafazadeh2} At this, most of the efforts were devoted to the so-called $\mathcal PT$-symmetric quasi-Hermitian evolution operators that have real spectrum. A real spectrum of its evolution operator makes a model a close relative with the conventional unitary quantum mechanics. This looks especially comforting when it comes to the physical interpretation of various constituents of the theory. In contrary, the spectra of Fokker-Planck operators are pseudo-Hermitian as they contain also complex conjugate pairs of eigenvalues. At this, there should be no concern about relevance of the non-real spectrum to reality - complex conjugate eigenvalues of transfer operators are well known in the DS theory under the name of Ruelle-Pollicott resonances. Thus, the STS is an "interpretable" theory with pseudo-Hermitian evolution operators and with applicability ranging from social sciences to astrophysics. This is particularly interesting in the light of the modern search for the physical realizations of pseudo-Hermitian evolution operators (see, \emph{e.g.}, Ref.\cite{OpticalREalization} and Refs. therein).

From a more general point of view, the classical DS theory and the supersymmetic and/or cohomological field theories (ChTs) are naturally synergetic within the STS. This synergy has a potential of a fruitful cross-fertilization between concepts and developments of these two major theoretical constructions. It is mainly in this line of thinking that in this paper the transfer operator (TO) formalism of the DS theory is given the STS representation. This approach can be looked upon as a bottom-up derivation of the STS with the starting point being the DS theory concept of the TO. This is opposed to the previously used Parisi-Sourlas stochastic quantization, \cite{ParisiSourlas} which is essentially a field theoretic construction rather than an intrinsic part of the modern theory of DSs. This alternative approach to the STS results in several additional links between the classical DS theory and the ChTs that lead to a few novel findings.

The main results of the paper are as follows. First, it provides a firm evidence that the spontaneous breakdown of topological supersymmetry is the field-theoretic essence and stochastic generalization of the concept of deterministic dynamical chaos. Second, it demonstrates that due to the unconditional existence of topological supersymmetry, the sharp trace and sharp determinant of the generalized transfer operator of the DS theory are subject to the supersymmetric trivialization. Third, it proves that Witten index of SDEs with closed phase spaces equals the Euler characteristic of the phase space for any smooth enough flow vector fields, noise temperatures, and noise metrics. In addition, it is demonstrated how the butterfly effect follows from the topological supersymmetry breaking picture of chaotic dynamics, and an unambiguous resolution of the Ito-Stratonovich dilemma is presented.

The structure of the paper is as follows. In Sec.\ref{TOSec}, the formalism of the Ruelle-Frobenuis-Perron TO is introduced. In Sec.\ref{TFTFromTO}, the supersymmetric pathintegral representation of the weighted traces of the TO is derived. In Sec.\ref{OperatorRep}, the operator representation of the theory is briefly discussed. In particular, it is shown that for closed phase spaces the Witten index always equals the Euler characteristic of the phase space. In Sec.\ref{GTO}, it is demonstrated that the generalized (stochastic) TO (GTO) formalism of the DS theory is an intrinsic part of the STS. In Sec. \ref{Discussion}, the supersymmetric trivialization of the sharp traces and determinants of the GTO is discussed. The phenomenon of the spontaneous breakdown of the topological supersymmetry is revisited and the situation is analyzed when the ground state's attenuation rate is negative. It is shown that in such situations the stochastically averaged number of periodic solutions/orbits grows exponentially in the large time limit. This is a direct indication on that the spontaneous breakdown of topological supersymmetry is the field-theoretic essence of (stochastic) chaotic dynamical behavior. It is also shown how the butterfly effect is derived from the topological supersymmetry breaking picture of chaotic dynamics. It is also argued that the topological supersymmetry must not be broken in the high temperature limit. This allows to extend the previously proposed low-temperature phase diagram (thermodynamic equilibrium / noise-induced chaos / ordinary chaos) onto the high temperature regime. Sec.\ref{Conclusion} concludes the paper. In the Appendix, a derivation of the Fokker-Planck evolution operator and the resolution of the Ito-Stratonovich dilemma in favor of the Weyl-Stratonovich approach are elaborated.
\newpage

\section{Transfer operator}
\label{TOSec}
We begin with the introduction of a continuous-time deterministic DS defined by the ordinary differential equation:
\begin{eqnarray}
\partial_t {x}(t) &=& {F}({x}(t)).\label{ODE}
\end{eqnarray}
Here, ${x} \in X$ are the DSs variables from the phase space, $X$, which is a $D$-dimensional topological manifold, and ${F}({x})\in T_{x}X$, is the flow vector field from the tangent space of $X$ at point $x$.

In order to avoid unnecessary complications of purely mathematical origin that go beyond the scope of this paper, we make several assumptions. First, the flow vector field is assumed to be continuous and smooth enough not to bring about possible complications associated with, \emph{e.g.}, discontinuous flows. Second, we assume that $X$ is orientable and has no boundary. Third, when it comes to the quantitative discussion of the Witten and Lefschetz indices, $X$ is also assumed compact, \emph{i.e.}, $X$ is closed.

Eq.(\ref{ODE}) defines a one-parameter group of diffeomorphisms of $X$, $ M_t(x): X \times \mathbb{R}\to X$, such that
\begin{eqnarray}
\partial_t  M_t(x) =  F(M_t(x)),\label{GroupOfDiffeomorohism}
\end{eqnarray}
and $M_0 = {Id}_{X}$. The Ruelle-Frobenius-Perron TO for the evolution of duration $t$ has the following form:
\begin{eqnarray}
(\mathcal{M}_{t}\rho)({x}) &=& \int \mathcal{M}_t(x,x')\rho({x}') d^D x' = \sum_{x', M_t(x')=x}\frac{\rho(x')}{| {det} TM_t(x')|},\label{RFPOT1}\\
\mathcal{M}_t(x,x') &=& \delta^D ({x}-{M}_t({x}')).\label{RFPOT}
\end{eqnarray}
where $|\cdot |$ stands for the absolute value, and
\begin{eqnarray}
TM_{t}{}^{i}_j(x) = \partial M_t^i( x)/\partial x^j,\label{DiffM}
\end{eqnarray}
is the coordinate representation of the tangent map induced by $M_t$:
\begin{eqnarray}
TM_{t}(x): T_xX \to T_{M_t(x)} X.\label{PushForward}
\end{eqnarray}
The flow is a group of diffeomorphisms and $M_t$ is invertible (for finite $t$), $M_t^{-1}=M_{-t}$, and any image, $x$, has only one preimage, $x'=M_{-t}(x)$, so that the summation sign in Eq.(\ref{RFPOT1}) is not really necessary. The generalization of this discussion onto the discrete-time DSs with non-invertable maps is beyond the scope of this paper.

Matrix (\ref{DiffM}) satisfies the following equation:
\begin{eqnarray}
\partial_t TM_t(x) &=& \hat {\mathbf  F}(M_t(x)) TM_t(  x),\label{TansferJacobian}\\
\hat {\mathbf F}(x) &\equiv & {\mathbf F}{}^{i}_j(x) = \partial {F}^i(x)/\partial x^j,\label{CurlyF}
\end{eqnarray}
as it follows from Eq.(\ref{GroupOfDiffeomorohism}). The formal solution of this equation with the initial condition $TM_0 = \hat 1_{TX}$ is
\begin{eqnarray}
TM_t(x) = :e^{\int_0^t \hat {\mathbf F}({M}_{t'}(x))dt'}:,
\end{eqnarray}
where columns denote chronological ordering. Also
\begin{eqnarray}
{det }TM_t(x) = e^{\int_{0}^t {Tr}\hat {\mathbf F}(M_{t'}(x))dt'}> 0,\label{PositivenessOfDet}
\end{eqnarray}
\emph{i.e.}, the flow preserves the orientation on $X$. The group structure of the flow is seen in
\label{Group}
\begin{eqnarray}
M_{t_1}(M_{t_2}(x))&=& M_{t_1+t_2}(x),\label{Group:U}\\
\mathcal{M}_{t_1}\mathcal{M}_{t_2}\rho &=& \mathcal{M}_{t_1+t_2}\rho,\label{Group:T}\\
TM_{t_1}(M_{t_2}(x))TM_{t_2}(x) &=& TM_{t_1+t_2}(x).\label{Group:Uc}
\end{eqnarray}
Let us define the weighted traces of the TO as
\begin{eqnarray}
{Tr}\mathcal{M}_{t} \phi
&=& \int \mathcal{M}_t(x,x)(\det TM_t(x))\phi(x)d^D  x =
\int \mathcal{M}_{-t}(x,x) \phi(x)d^D x
\nonumber\\&=&
\sum_{x\in{fix} M_t}\frac{\phi(x)}{| {det}(\hat 1 - TM_{-t}(x) )|}.
\label{WeightedTrace}
\end{eqnarray}
Here $TM_{-t}(x) = TM_{t}^{-1}(x)$, Eq.(\ref{PositivenessOfDet}) is utilized, and $\phi(x)$ is some function on $X$ called weight function. Eq.(\ref{WeightedTrace}) has a seeming appearance of time-reversed evolution. Its meaning will be discussed in Sec. \ref{GTOPathintegral}. The sum notation in Eq.(\ref{WeightedTrace}) suggests that the fixed points of $M_t$ are isolated. The pathintegral version of the theory, however, deals equally well with more general situations with non-isolated fixed points as we briefly discuss at the end of this section. Having this in mind, we will sloppily keep  using the summation sign.

The following two weight functions are of specific interest:
\label{weights}
\begin{eqnarray}
w(x) &=& {det} (\hat 1 - TM_{-t}(  x)) = \sum\nolimits_{k=1}^D (-1)^k m_k(x),\label{weights:W}\\
z(x) &=& {det} (\hat 1 + TM_{-t}(  x)) = \sum\nolimits_{k=1}^D m_k(x),\label{weights:Z}
\end{eqnarray}
where
\begin{eqnarray}
&m_k(x) = \sum\limits_{i_1<...<i_k} {det} \left( \begin{array}{ccc}
TM_{-t}{}^{i_1}_{i_1}(  x)& \cdots & TM_{-t}{}_{i_1}^{i_k}(x) \\ \vdots & \ddots & \vdots\\ TM_{-t}{}_{i_k}^{i_1}(  x)& \cdots & TM_{-t}{}_{i_k}^{i_k}(  x)\end{array}\right), \label{weights:FP}
\end{eqnarray}
and the characteristic polynomial formula was used
\begin{eqnarray}
{det}(\hat 1 + \lambda TM_{-t}(  x)) = \sum\nolimits_{k=0}^{D}\lambda^k m_k(x). \nonumber
\end{eqnarray}

The weighted trace with weight function (\ref{weights:W})
\begin{eqnarray}
{Tr}\mathcal{M}_{t}w &=& \sum_{  x\in{fix} M_t}{sign } {det}(\hat 1- TM_{-t}(  x) ),\label{LefschetzHoft}
\end{eqnarray}
is of topological origin for it is the matter of the Lefschetz-Hopf theorem. The later states, in particular, that for closed phase spaces,
\begin{eqnarray}
{Tr}\mathcal{M}_{t}w &=& \sum\nolimits_{k=0}^D (-1)^{k} {Tr}_{H^k} M_{-t}^*,\label{LefschetzHoft1}
\end{eqnarray}
where ${Tr}_{H^k}$ denotes the trace over the $k^{th}$-degree cohomology of $X$ and $M_{-t}^*$ is the pullback induced by $M_{-t}$. The rhs of Eq.(\ref{LefschetzHoft1}) is called Lefschetz number or index. In the $t\to0$ limit, $M_{-t}\to{Id}_X$ and the summation in the r.h.s. is the summation over the signed Betti numbers so that the index evaluates to the Euler characteristic of $X$.

In fact, the topological nature of Eq.(\ref{LefschetzHoft}) will be reassured later by its identification with the Witten index, $W$, of the corresponding STS:
\begin{eqnarray}
W_{cl} = {Tr}\mathcal{M}_{t} w.
\end{eqnarray}
(Here and in the following the subscript '\texttt{cl}' points onto the deterministic or "classical" nature of an object as opposed to the stochastic objects with no subscripts.) The identification of Eq.(\ref{LefschetzHoft}) as the Witten index immediately suggests that it is independent of $t$ (see below). The physical explanation of the topological nature of the Witten index is that it is the representative (up to a topological factor) of the normalized partition function of the stochastic noise,\cite{Mine3} which is independent of all the details of the model.

The meaning of the other weight function is best seen in the opposite limit of the infinitely long temporal evolution, $t\to\infty$. Eq.(\ref{WeightedTrace}) can be rewritten as
\begin{eqnarray}
{Tr}\mathcal{M}_{t} z = \sum_{x\in{fix}  M_t}\frac{{det}(TM_{t}(x) + \hat 1)}{|{det}(TM_{t}(x) - \hat 1)|}.\label{ZTRace}
\end{eqnarray}
Matrix $TM_{t}(x)$ is real and its eigenvalues can be expressed as $e^{-\lambda_a(t) t}, a=1,...,D$, where $\lambda(t)$'s are either real or come in complex conjugate pairs. By the argumentation that stands behind the introduction of global Lyapunov exponents, one assumes that there exist a class of DSs with well defined limits, $\lim_{t\to\infty}\lambda_a(t)=\lambda_a$. If none of ${Re}\lambda_a$ is exactly zero,
\begin{eqnarray}
\lim_{t\to\infty}\frac{{det}(TM_{t}(x) + \hat 1)}{|{det}(TM_{t}(x) - \hat 1)|} = \lim_{t\to\infty}\prod\nolimits_{i=a}^D\frac{e^{-\lambda_a t}+1}{|e^{-\lambda_a t}-1|} = 1.
\end{eqnarray}
Therefore, for this class of models and in the long-time limit, Eq.(\ref{ZTRace}) is the number of fixed points of $M_t$. These fixed points include the fixed points of the flow, $F$, as well as periodic solutions/orbits that must contribute when the duration of time is a multiple of their periods. In other words, the number of the fixed points of $M_t$ can be interpreted as the number of periodic solutions with the fixed points of the flow vector field being the trivial (constant in time) realization of this concept.

Even in those situations when Eq.(\ref{ZTRace}) does not represent the number of periodic solutions, it still must be interpreted as the physical or dynamic partition function of the SDE. As it will be discussed in Secs.\ref{GTO} and \ref{OperatorRep}, Eq.(\ref{ZTRace}) is the trace of the most fundamental object of the theory - the stochastically generalized TO and/or the finite-time Fokker-Planck evolution operator. This identification justifies the introduction of the standard notation for the partition function:
\begin{eqnarray}
Z_{cl} = {Tr}\mathcal{M}_{t} z.\label{ZClassical}
\end{eqnarray}
We would also like to point out that fixed points of $M_{t}$ that come from periodic solutions are not isolated, \emph{i.e.}, every point of the closed trajectory is a fixed point of $M_{t}$ when the time duration is a multiple of the period. Furthermore, the fixed points of the flow vector field (which are also the fixed points of $M_{t}$) may also be non-isolated on their own and form submanifolds of $X$ (\emph{e.g.}, the Bott-Morse case). How to count non-isolated fixed points is not very clear within the methodology in this section. In particular, on a non-isolated fixed point, the matrix $TM_{t}(x)$ has at least one eigenvalue which is exactly unity so that the determinant in the denominator in Eq.(\ref{ZTRace}) is zero and the sign in Eq.(\ref{LefschetzHoft}) is not well defined.

This emphasizes once again the advantage of the pathintegral/operator representation of the theory. There, such problem does not exist as the theory deals equally well with the non-isolated fixed points of $M_t$. For example, in Bott-Morse case, where the Langevin stochastic dynamics is defined by a flow which is a gradient of some function with critical points forming submanifolds of the phase space, every cohomology class of these invariant manifolds provides a perturbative or local supersymmetric ground state. In result, the perturbative-level contribution from each invariant manifold into the Witten index is (up to a sign) its Euler characteristic. \cite{Labastida}

\section{Pathintegral representation}
\label{TFTFromTO}
\begin{figure}[t]
\begin{center}
\includegraphics[width=10cm,height=4.6cm]{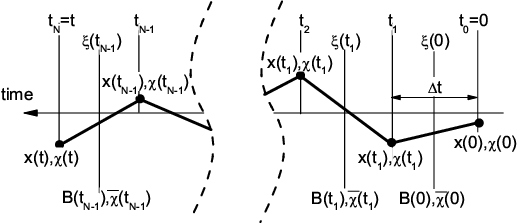}
\end{center}
\caption{\label{Fig1} The fields of the pathintegral representation of the theory. The entire interval of temporal evolution is divided into $N\to\infty$ segments. At each time slice, $t_p, 0\le p \le N$, one has the field, $x(t_p)\in X(t_p)$, from a copy of the phase space, $X(t_p)$, and the anticommuting ghost field, $\chi(t_p)\in T_{x(t_p)}X(t_p)$, from the tangent space of $X(t_p)$. In between the time-slices, there are the Langrange multiplier and the antighost, $B(t_p),\bar\chi(t_p)\in T^*_{x(t_p)}X(t_p)$, both from the cotangent space of $X(t_p)$ (or $X(t_{p+1})$). $B(t_p)$ and $\bar\chi(t_p)$ are needed for the exponentiation of the $\delta$-functions representing the infinitesimal temporal evolution of the bosonic fields and the ghosts. In case of stochastic models, there are also variables representing the noise, $\xi(t_p)$. Integration of all the fields with the periodic boundary conditions, $x(t)=x(0), \chi(t)=\chi(0)$, leads to the Witten index. The integration with the antiperiodic boundary conditions for the ghosts, $x(t)=x(0), \chi(t)=-\chi(0)$, gives the dynamical  partition function. Integrating out only the intermediate fields and leaving the initial, $x(0),\chi(0)$, and final, $x(t), \chi(t)$, fields unspecified, gives the finite-time Fokker-Planck evolution operator known in the DS theory as the generalized transfer operator.}
\end{figure}

In this section, the TO formalism is given a pathintegral representation. The fields involved in this formulation are schematically presented in Fig.\ref{Fig1}.

\subsection{Commuting fields}
With the help of Eq.(\ref{Group:T}), Eq.(\ref{WeightedTrace}) can be rewritten as
\begin{eqnarray*}
&{Tr}\mathcal{M}_{t} \phi = \int \phi(x(0)) \prod\limits_{p=0}^{N-1} \mathcal{L}_{ -\Delta t}(x(t_{p}), x(t_{p+1})) d^D x(t_p).
\end{eqnarray*}
Here the time domain $t\in(0,t)$ is split into $N$ segments $(t_{p-1},t_{p})$, with $t_p = p t / N$ being equally separated time slices, $\Delta t = t / N$, $x(t_p)$ are the intermediate variables at the corresponding times, and the periodic boundary conditions are assumed, $x(t_N) \equiv x(t)=x(0)$.

Now we introduce an additional field $B(t_p)\in T_{x(t_p)}^*X$ called Lagrange multiplier that belongs to the cotangent space of $X$, in order to exponentiate the $\delta$-functions:
\begin{eqnarray}
\mathcal{M}_{-\Delta t}(x(t_{p}), x(t_{p+1})) = \delta^D\left(x(t_{p})
- M_{-\Delta t}(x(t_{p+1})) \right)\nonumber \\ = \int e^{i B_i(t_p) \left(M^i_{-\Delta t}(x(t_{p+1})) -
x^i(t_{p})\right) } d^D \left(\frac{B(t_p)}{2\pi}\right).\nonumber
\end{eqnarray}
Taking now the continuous-time limit ($N\to\infty$), one finds that
\begin{eqnarray}
{Tr}\mathcal{M}_{t} \phi &=& \int \phi(x(0) ) e^{ S_{B} }D   x D   B, \label{Trace:Bosonic}
\end{eqnarray}
where the path integration is over the closed trajectories (periodic boundary conditions), the action is:
\begin{eqnarray}
S_{B}(x,B) &=& \lim_{N\to\infty}i \sum_{p=0}^{N-1} B_i(t_p) \times (M^i_{-\Delta t}(  x(t_{p+1})) - x^i(t_{p}))\nonumber \\
&=& i \int dt B_i(t) \left(\partial_t x^i(t) - F^i (x(t))\right),\label{BosonAction}
\end{eqnarray}
and the differential of the pathintegral is
\begin{eqnarray}
Dx DB = \lim_{N\to \infty} \prod\nolimits_{p=0}^{N-1} d^Dx(t_p) d^D\left(\frac{B(t_p)}{2\pi}\right).\label{DxDB}
\end{eqnarray}

\subsection{Anticommuting fields}
Expression (\ref{Trace:Bosonic}) has the form of a pathintegral expectation value of some temporarily local observable. This, however, is not quite correct because the weight functions of interest in Eq.(\ref{weights}) are temporarily nonlocal function(al)s of $x(t)$. A consistent pathintegral representation of these weight functions can be achieved with the help of anticommuting fields called Fadeev-Popov ghosts, $\chi^i, \bar\chi_i, i=1...D$. The ghosts obey Berezin rules of integration:\cite{CoherentFermionicStates}
\begin{eqnarray}
\int \chi^i d\chi^j = - \int d\chi^j \chi^i = \delta^{ij}, \int d\chi^i = 0,
\end{eqnarray}
(and similar relation for the anti-ghosts $\bar\chi_i$ introduced below). These rules suggest in particular the following useful properties:
\begin{eqnarray}
\int d^D \bar{  \chi} e^{\bar\chi_i \hat A^i_j \chi^j} &=& \delta^D \left(\hat A\chi\right) ,\label{FermPRop:DeltaExp}\\
\int d^D\chi \delta^D \left(\hat A\chi\right) &=& {det}\hat A.\label{FermPRop:Det}
\end{eqnarray}
Here $d^D \bar{  \chi} = d \bar{  \chi}^D...d \bar{  \chi}^1$, $d^D \chi= d   \chi^1...d \chi^D$, and $\delta^D$ is the "fermionic" $\delta$-function, $\delta^D(\chi) = (-1)^{D}\chi^D...\chi^1$, that behaves on integration similarly to its bosonic counterpart:
\begin{eqnarray}
\int d^D\chi \delta^D(\chi-\xi)f(\chi) = f(\xi),
\end{eqnarray}
where $f(\chi)$ is any function of the anticommuting field.

With the help of Eqs.(\ref{FermPRop:Det}), the weight function from Eq.(\ref{weights:W}) can be given as
\begin{eqnarray}
w(  x(0)) = {det}\left(\hat 1 - TM_{-t}(x(0)) \right) =\int_{PBC} d^D \chi(0)\delta^D\Bigl(\chi(0) - \hat TM_{-t}(x(0))\chi(t)\Bigr),\label{Pathphi:W}
\end{eqnarray}
where $\chi (0), \chi (t)\in T_{  x(t)}X$ are the anticommuting ghosts at the corresponding times and the subscript $PBC$ signifies periodic boundary conditions for the ghosts: $\chi (t) \equiv \chi (0)$. Using Eq.(\ref{Group:Uc}) one can now utilize again the time slices' picture of the previous subsection and bring Eq.(\ref{Pathphi:W}) to the following form:
\begin{eqnarray}
w(x(0)) = \int_{PBC} \prod_{p=0}^{N-1}d^D \chi(t_p)\delta^D\Bigl(\chi(t_p) - TM_{-\Delta t}(  x(t_{p}))\chi(t_{p+1})\Bigr) .
\end{eqnarray}
Just as in the case of bosonic fields, one can further exponentiate the integrand in the previous expression with the help of Eq.(\ref{FermPRop:DeltaExp}) and by the introduction of yet another anticommuting ghost field from the cotangent space of $X$, $\bar{\chi}(t_p)\in T^*_{x(t_p)}X$. In the continuous time limit, $N\to\infty$, one arrives at
\begin{eqnarray}
w(x(0)) &=& \int_{PBC}e^{S_{F}} D\chi D\bar{ \chi} ,
\end{eqnarray}
where
\begin{eqnarray}
S_{F}(x,\chi,\bar\chi) &=& \nonumber i \lim_{N\to\infty}\sum_{p=0}^{N-1}\bar\chi_i(t_p) \times \Bigl( \chi^i(t_p) - TM_{-\Delta t}{}^i_j(  x(t_{p}))\chi^j(t_{p+1})\Bigr)\nonumber\\
&=&-i \int dt \bar{\chi}_i(t) \left(\partial_t \chi^i(t) - \hat{\mathbf F}^i_j (  x(t)) \chi^j(t)\right),\label{FermiAction}
\end{eqnarray}
with $\hat {\mathbf F}$ from Eq.(\ref{CurlyF}), and the integration measure being
\begin{eqnarray}
D  \chi D\bar{ \chi} = \lim_{N\to \infty} \prod\nolimits_{p=0}^{N-1} d^D   \chi(t_p)d^D (i \bar{  \chi}(t_p)). \label{chibarchi}
\end{eqnarray}
The factor $i$ is not necessarily and it can be absorbed by the redefinition of the antighost field, $\bar\chi$. The reason for its introduction is to bring the operator of $\mathcal Q$ symmetry below to its conventional form.

\subsection{Emergence of Topological Supersymmetry}
Combining Eqs. (\ref{BosonAction}) and (\ref{FermiAction}), one finds
\begin{eqnarray}
W_{cl} = \int_{PBC} e^{S_{cl}(\Phi)} D\Phi,\label{TFT}
\end{eqnarray}
where $\Phi = (x,B,\chi,\bar{\chi})$ is the collection of all the fields, the total functional differential is $D\Phi = Dx DB D\chi D\bar\chi$ defined in Eqs.(\ref{DxDB}) and (\ref{chibarchi}), and the total action, $S_{cl} = S_B + S_F$, is the sum of the bosonic and fermionic parts from Eqs.(\ref{BosonAction}) and (\ref{FermiAction}).

As it can be straightforwardly verified, the action of the theory is $\mathcal Q$-exact, \emph{i.e.}, it can be represented as
\begin{eqnarray}
S_{cl}(\Phi) = \{\mathcal{Q}, \Psi_{cl}(\Phi)\}, \label{Action}
\end{eqnarray}
where $\mathcal Q$ is the nilpotent operator, that is $\{\mathcal{Q}, \{ \mathcal{Q}, X(\Phi) \}\}=0, \forall X(\Phi)$, of the topological supersymmetry and/or Becchi-Rouet-Stora-Tyutin symmetry
\begin{eqnarray*}
\{\mathcal{Q},X(\Phi)\} = \int dt\left(\chi^i(t) \frac{\delta}{\delta x^i(t)} + B_i(t) \frac{\delta}{\delta \bar \chi_i (t)}\right) X(\Phi),
\end{eqnarray*}
acting on the so-called gauge fermion,
\begin{eqnarray}
\Psi_{cl}(\Phi) = i \int dt \bar \chi_i(t)\left(\partial_t x^i(t) - F^i(  x(t))\right).\label{GaugeFermion}
\end{eqnarray}
The wideast class of models with $\mathcal Q$-exact actions, such as the one in Eq.(\ref{TFT}), are the cohomological field theories. \cite{TFT,WittenTFT1,WittenTFT2,Labastida,Frenkel,Blau,Atiyah,Anselmi} Such theories allow for the calculation of certain topological invariants as expectation values of $\mathcal Q$-closed operators on supersymmetric states. We will discuss this issue in some more details in Sec.\ref{BeyondInstantons} below.

\subsection{Generalization to non-autonomous DS}
\label{SecNonAutomatous}
The above derivations can be generalized to cases of nonautonomous DSs, \emph{i.e.}, to DSs with time dependent flow fields. Indeed, let us note that the derivations that led to from Eq.(\ref{WeightedTrace}) to Eq.(\ref{TFT}) did not rely on the assumption that the flow vector field has no explicit dependence on time. Thus, these steps can be repeated also for time dependent flows, $F(x,t)$. The only difference this will lead to is the substitution $F(x(t))\to F(x(t),t)$ in the definition of the gauge fermion in Eq.(\ref{GaugeFermion}).

\subsection{Stochastic generalization}
\label{SecStoch}
Moreover, it is possible to generalize further to the case of stochastic external influence, \emph{i.e.}, the noise. First, let us isolate the time dependent part of the flow in the following manner:
\begin{eqnarray}
F^i(x,\xi(t)) = F^i(x) + (2\Theta)^{1/2} e^i_a(x)\xi^a(t).\label{Fmodified}
\end{eqnarray}
Here functions $e^i_a(x)$ can be interpreted as vielbeins, $\xi^a(t)$ are parameters representing external influence, and $\Theta$ is its temperature/intensity.\footnote{The definition of temperature here is conventional for the literature on SDEs, and it differs by the factor of 2 from that was used in Ref.\cite{Mine3}.} In the literature on stochastic DSs, the situations with coordinate-dependent/independent vielbeins are often called multiplicative/additive noise.

One considers now the stochastic expectation value of the Witten index:
\begin{eqnarray}
W = \langle W_{cl}(\xi)\rangle_{Ns} \equiv \int W_{cl}(\xi)P(\xi) D\xi, \label{StochW}
\end{eqnarray}
where $P(\xi)$ is the normalized probability density functional of the configurations of the noise. $P(\xi)$ can be of a very general form. For example, the noise can be nonlocal in time, nonlinear, and it can as well have a non-vanishing "classical" component, $\langle \xi(t)\rangle_{Ns}\ne0$. In Eq.(\ref{StochW}),
\begin{eqnarray}
W_{cl}(\xi) = \int_{PBC} e^{S_{cl}(\Phi,\xi)}D\Psi,\label{TFTXi}
\end{eqnarray}
with the action $S_{cl}(\Phi,\xi) = \{\mathcal{Q}, \Psi_{cl}(\Phi,\xi)\}$ defined by the gauge fermion
\begin{eqnarray}
\Psi_{cl}(\Phi,\xi) = i \int dt \bar \chi_i(t)\left(\partial_t x^i(t) - F^i(x(t),\xi(t))\right).\label{GaugeFermionXi}
\end{eqnarray}

Integrating out $\xi$'s, one arrives at the new action
\begin{eqnarray}
S(\Phi) &=& S_{cl}(\Phi) + \Delta S(y),\label{newaction}
\end{eqnarray}
where $S$ is from Eq.(\ref{Action}), while the part provided by the noise is
\begin{eqnarray}
\Delta S(y) &=& \log \langle e^{\int_t y_a(t)\xi^a(t)} \rangle_{Ns} \nonumber \\
&=& \sum_{k=1}^\infty \int dt_1...dt_k (k!)^{-1} C_{(k)}^{a_1..a_k}(t_1...t_k)\prod_{j=1}^k y_{a_j}(t_j),\label{newactiondelt}
\end{eqnarray}
with $C$'s being the (irreducible) correlators of the noise and
\begin{eqnarray}
y_a(t) = \{\mathcal{Q}, -i (2T)^{1/2}\bar\chi_i(t) e^i_a(x(t)) \}.\label{ycoupling}
\end{eqnarray}
Due to the nilpotency of the differentiation by $\mathcal Q$, \emph{i.e.}, $\{\mathcal Q,\{\mathcal{Q},\Phi\}\} = 0$, the product of any number of $\mathcal Q$-exact factors is $\mathcal Q$-exact itself, $\{\mathcal Q,X_1\}  \{\mathcal Q,X_2\}... = \{\mathcal Q, X_1 \{\mathcal Q, X_2\}...\}$. Therefore, $\Delta S$ in Eq.(\ref{newactiondelt}), which is a functional only of $\mathcal Q$-exact $y(t)$'s, is $\mathcal Q$-exact together with the entire new action
\begin{eqnarray}
S = \{\mathcal{Q},\Psi(\Phi)\},
\end{eqnarray}
with the new gauge fermion being
\begin{eqnarray}
\Psi(\Phi) &=& \Psi_{cl}(\Phi) + \sum_{k=1}^\infty \int dt_1...dt_k (2T)^{k/2} (k!)^{-1} C_{(k)}^{a_1..a_k}(t_1...t_k) \nonumber \\ &&\times (-i \bar\chi_{i_1}(t_1)) e_{a_1}^{i_1}(  x(t_1))\prod_{j=2}^k \{\mathcal{Q}, -i \bar\chi_{i_j}(t_j) e_{a_j}^{i_j}(  x(t_j)) \},
\end{eqnarray}
and with $\Psi_{cl}(\Phi)$ from Eq.(\ref{GaugeFermion}). Therefore, after the generalization to non-autonomous and/or stochastic DSs, the model is still of topological nature:
\begin{eqnarray}
W = \int_{PBC} e^{S(\Phi)}D\Phi.\label{TFTStoch}
\end{eqnarray}

\subsection{Gaussian white noise}

For the Gaussian white noise, for which $C^{ab}_{(2)}(t_1t_2)=\delta^{ab}\delta(t_1-t_2)$ and all the other $C$'s vanish, the gauge fermion has the following form:
\begin{eqnarray}
\Phi(\Phi) &=& \int dt (i\bar\chi_i \partial_t x^i - j(\Phi(t))),\\
j(\Phi(t)) &=& i\bar\chi_i \left(F^i(x) -  T g^{ij}(iB_j -\tilde\Gamma_{kj}^l\chi^k(i\bar\chi_l))\right),\label{GaugeCurrent}
\end{eqnarray}
where $g_{ij}=(g^{ij})^{-1}=e^a_ie_j^a$ is the "noise-induced" metric and
\begin{eqnarray}
\tilde\Gamma_{ji}^k= - e^a_i\partial_{j}e^{k}_a = e^{k}_a\partial_{j}e^a_i,\label{WeitzenbockConnection}
\end{eqnarray}
is the Weitzenb\"ock connection that has a non-vanishing torsion and zero curvature tensor (see, e.g., Ref.\cite{Weitzenbock}). $\tilde\Gamma$ is compatible with the metric:
\begin{eqnarray}
\tilde \nabla_ig^{jk}=\partial_i g^{jk} + \tilde \Gamma_{ip}^jg^{pk} + \tilde \Gamma_{ip}^kg^{pj}=0.\label{MetricCompat}
\end{eqnarray}

\subsection{Dynamic partition function}
\label{PhysPartFunc}
The trace of TO with the weight function from Eqs.(\ref{weights:Z}) also admits a pathintegral representation similar to Eq.(\ref{TFTStoch}). Using again (\ref{FermPRop:Det}), the weight function from Eq.(\ref{weights:Z}) can be represented as
\begin{eqnarray}
z(x(0)) &=& {det}\left(\hat 1 + TM_{-t}(x(0))\right) \nonumber \\&=&\int_{APBC} \delta^D\Bigl(\chi(0) - TM_{-t}(  x(0))  \chi(t)\Bigr)d^D   \chi(t),\label{Pathphi:Z}
\end{eqnarray}
where the subscript indicates antiperiodic boundary conditions: $\chi(t)=-\chi(0)$. This is the analogue of Eq.(\ref{Pathphi:W}) with the only difference in the boundary conditions for the ghosts. This difference, however, does not interfere with the steps that led from Eq.(\ref{Pathphi:W}) to Eq.(\ref{TFTStoch}). Therefore:
\begin{eqnarray}
Z = \langle Z_{cl}(\xi)\rangle_{Ns} = \int_{APBC} e^{S(\Phi)}D\Phi,\label{TFTZ}
\end{eqnarray}
with $S(\Phi)$ defined in Eq.(\ref{newaction}) and with $Z_{cl}$ being the deterministic dynamic partition function defined in Eq.(\ref{ZClassical}). According to the discussion at the end of Sec.\ref{TOSec}, in the $t\to\infty$ limit and for some DSs, Eq.(\ref{TFTZ}) represents the stochastically averaged number of periodic solutions. It is also worth noiting that the antiperiodic boundary conditions for ghosts are not compatible with the $\mathcal Q$-symmetry so that $Z$ is not of topological character and in particular depends on many parameters of the model as well as on the duration of temporal propagation.    

\subsection{STS and Cohomological Theories}
\label{BeyondInstantons}

Finalizing the discussion in this section we would like to clarify here the following subtle point. Within the formulation of the cohomological field theories (see, \emph{e.g.}, \cite{TFT}), besides the requirement on the $\mathcal Q$-exact action one also restricts his attention to the expectation values of $\mathcal Q$-closed operators (topological or BPS observables, \cite{Frenkel}) on supersymmetric states. In STS, on the other hand, one is  interested primary in the ground states of the theory. In "chaotic" situations that are most intriguing from the point of view of richness of dynamics, the ground states of the model are non-supersymmetric. Therefore, the STS can not be identified as a full-flegded cohomological field theory.

Nevertheless, the STS and the cohomological field theories have very close connection. In particular, the Witten index (\ref{TFTStoch}) is of topological origin and it is actually a standard pathintegral representation of the corresponding cohomological theory. Furthermore, when the $\mathcal Q$-symmetry is not spontaneously broken and one is interested in the physical limit of the infinitely long temporal evolution, he can restrict his attention to only the ground states that are all supersymmetric. Studying the expectation values of collections of only $\mathcal Q$-closed operators on these states will turn the STS into a cohomological theory. In fact, the response operators in Sec.\ref{SecButterflyEffect} below are $\mathcal Q$-exact and thus are $\mathcal Q$-closed. In this case, however, the expectation values of these response operators vanish in case of unbroken $\mathcal Q$-symmetry thus providing no interesting information except that the model does not exhibit the spontaneous long-range memory/response, \emph{i.e.}, the butterfly effect. 

Moreover, in the cohomological theories one is often interested in calculating expectation values of $\mathcal Q$-closed operators on instantons (see, \emph{e.g.}, Ref.\cite{Anselmi}) - the classical solutions that connect invariant manifolds of different stability.  Instantons and the BPS observables on them may also be useful for certain purposes in the STS. For instance, instantons can be physically realized as quenches. Upon a quench, the parameters of a DS are abruptly changed and the DS suddenly finds itself away from its (new) stable attractors. In some realizations, this new initial position may as well lie on the boundary of the new attraction basins. 

One example of such a situation is a quench across a symmetry breaking phase transition. There, the initial position right after the quench is the state with the order parameter zero everywhere, while the new attractors (below the transition point) are various solitonic configurations of a non-zero order parameter. The quench evolution is a complicated instanton connecting the initial position and one of the stable attractors. The expectation values of $\mathcal Q$-closed operators may be useful in this situation, \emph{e.g.}, for the purpose of the classification of the attractors/solitonic configurations and perhaps even for the analysis of the probability to end up (after the quench) in one or another solitonic configuration.

Interestingly enough, from the results of Ref.\cite{Frenkel} it follows that the effective low-energy theory of any instanton/quench is a log-conformal field theory. In other words, even quenches that are not across a phase transition must exhibit the long-range chaotic order. A natural example of this behavior is, \emph{e.g.}, the Barkhausen effect or the crumpling paper that can be viewed as a "slow" quench and that exhibits the long-range chaotic order in the form of the algebraic statistics of observables related to the avalanches/crackles such as the "mass" of the avalanches or the waiting time between the avalanches. The emergence of the long-range chaotic order in the effective low-energy theories of instantons/quenches can be attributed to the intrinsic breakdown of the $\mathcal Q$-symmetry within any instanton.

Instantons on their own can not represent, however, the global ground state of a DS because instantons must be compensated by anti-instantons in order to implement the periodic boundary conditions in time. Nevertheless, the methodology of the BPS or topological observables on instantons may find its application even for the studies of the global ground states. One such possibility is the use of the instantoic matrix elements (in combination of with their anti-instantonic counterparts) for the construction of the effective low-energy theories of the noise-induced chaotic DSs, in which $\mathcal Q$-symmetry is broken by the condensation of (anti-)instantonic configurations that from the physical point of view are the noise-induced tunneling matrix elements between, \emph{e.g.}, different attractors (see Sec.\ref{GeneralPhaseDiagram}).

\section{Operator representation}
\label{OperatorRep}
One can now pass to the operator representation of the theory where $x$'s and $\chi$'s are diagonal, whereas
\begin{eqnarray}
i\hat {B} = \partial/\partial   x, i\hat{\bar {  \chi}} = \partial/\partial  \chi.\label{AdditionalFields}
\end{eqnarray}
The Taylor expansion in $\chi$'s of a wavefunction, $\psi(  x,  \chi)$, terminates at the $D$'th term:
\begin{eqnarray}
\psi(x,\chi) &=& \sum\nolimits_{k=0}^D \frac{1}{k!} \psi^{(k)}(x),\\\label{wavefunctionkth}
\psi^{(k)}(x) &=& \psi_{i_1...i_k}(x) \chi^{i_1} ... \chi^{i_k},
\end{eqnarray}
because all combinations $\chi^{i_1}...\chi^{i_k}$ with $k>D$ vanish due to the anticommutativity of ghosts.

For white noises (but not necessarily Gaussian noises), the time evolution of a wavefunction is given by the Fokker-Planck equation
\begin{eqnarray}
\partial_t \psi = -\hat H\psi.\label{FokkerPlanck}
\end{eqnarray}
The explicit form of the Fokker-Planck operator can be established in accordance with the rules of the bi-graded Weyl symmetrization, which is equivalent to the Stratonovich approach to SDEs, of its pathintegral expression in
\begin{eqnarray}
S = \int dt (i B_i\partial_t x^i - i\bar\chi_i\partial_t\chi^i- H(\Phi)),
\end{eqnarray}
where
\begin{eqnarray}
H(\Phi) = \left\{ {\mathcal Q}, j(\Phi) \right\},\label{PathIntegralH}
\end{eqnarray}
and for Gaussian white noise $j$ is given in Eq.(\ref{GaugeCurrent}). The Fokker-Planck operator is: (see Appendix for details of the derivation)
\begin{eqnarray}
\hat H = [\hat d, \hat j].\label{HamiltonianTFT}
\end{eqnarray}
Here $\hat d = \chi^i \partial/\partial x^i$ is the conserved N\"{o}ther charge associated with the $\mathcal Q$-symmetry so that
\begin{eqnarray}
[\hat d, \hat H] = 0.\label{CommutationHd}
\end{eqnarray}
The operator version of Eq.(\ref{GaugeCurrent}), $\hat j = \partial/\partial\chi^i F^i + \Theta \hat {\bar d}$, with $\hat {\bar d} = - \partial/\partial\chi^i g^{ij}\hat {\tilde \nabla}'_j$ and $\hat {\tilde \nabla}'_j = \partial/\partial x^j - \tilde\Gamma^l_{kj} \chi^k (\partial/ \partial\chi^l)$, is sometimes interpreted in the literature as the probability current.

In Eqs.(\ref{HamiltonianTFT}) and (\ref{CommutationHd}) we introduced the bi-graded commutator of operators. It is defined as an anti-commutator if both operators are fermionic, \emph{i.e.}, have odd total number of $\chi$'s and $\partial/\partial\chi$'s, and as a commutator otherwise so that in Eqs.(\ref{HamiltonianTFT}) and (\ref{CommutationHd}) it is respectively the anti-commutator and commutator.

Note also that the Fokker-Planck operator (\ref{HamiltonianTFT}) can not be recognized as N=2 supersymmetric as in Langevin SDEs because in general $\hat j^2\ne0$ and Eq.(\ref{HamiltonianTFT}) can not be represented as a square of an operator with mixed ghost degree. Neither can it be given as a square of a fermionic operator as in case of Kramers Eq. \cite{Kramers} In a general case, a $\hat d$-exact evolution operator can only be recognized as that of a topological quantum mechanics \cite{Labastida} or as the $(1,0)$ supersymmetry.

The Fokker-Planck operator is a real but not Hermitian operator, $\hat H\ne\hat H^\dagger$. Its eigenvalues are either real or come in complex conjugate pairs, whose DS theory counterparts are known as Ruelle-Pollicott (RP) resonances:
\begin{eqnarray}
\mathcal{E}_r = \Gamma_r, \mathcal{E}^\pm_p = \Gamma_p \pm i E_p.\label{Eigenvalues}
\end{eqnarray}
Here, the real parts of the eigenvalues, $\Gamma$'s, are the attenuation rates or inverse lifetimes of the eigenstates, while the imaginary parts, $E$'s, can be looked upon as energies in the quantum mechanical sense.

\begin{figure}[th]
\begin{center}
\includegraphics[width=10cm,height=6cm]{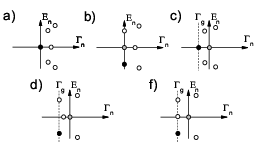}
\end{center}
\caption{\label{Fig2} Various Fokker-Planck spectra in relation to the phenomenon of the spontaneous breakdown of $Q$-symmetry. Among all the eigenstates with $\Gamma_n = \min_{n}\Gamma_n\equiv \Gamma_g$, the ground state (indicated as black circles) is the one with the lowest $E_n$. There always exist at least one $\mathcal Q$-symmetric "thermodynamic equilibrium" (TE) eigenstate (gray circle at the origin, see discussion in Sec.\ref{Discussion}). This state is among the ground states only when $\mathcal Q$-symmetry is not broken spontaneously {\bf (a)} so that after infinitely long temporal evolution the DSs is always at its TE. Figs. {\bf (a)} and {\bf (b)} correspond to cases when $\Gamma_g=0$. For all except (a) the topological supersymmetry is broken. The $\eta$T-time reversal symmetry (see discussion after Eq.(\ref{Eigenvalues})) is broken for {\bf (b)}, {\bf (d)}, {\bf (f)}. Some forms of spectra may not be realizable, however, as is discussed in Sec.\ref{SecChaos}.}
\end{figure}

The ground state can be uniquely defined (up to the $\mathcal Q$-symmetry degeneracy) for each spectrum in Fig.\ref{Fig2} with the help of two arguments. The first one is straightforward: the ground state must have the lowest possible attenuation rate, $\Gamma_n$, so that only the ground states survive a sufficiently long temporal Fokker-Planck evolution of any wavefunction, whereas all the other eigenstates will be exponentially suppressed. The second argument is more subtle. If there are more than one such eigenstates with different imaginary parts of the eigenvalues, $E_n's$, than the ground state is the one with the lowest $E_n$. The same argument applies to the unitary quantum mechanical models, in which in our terms all $\Gamma$'s are zero. One of the justifications for this argument is the possibility to Wick rotate time a little ($t \to t+0^+$) so that the so-chosen ground state would provide the dominant contribution into the dynamic partition function in the long-time limit.

An operator with this form of spectrum can be recognized as pseudo-Hermitian. \cite{Mostafazadeh1,Mostafazadeh2} In particular, it must possess the so-called $\eta T$-symmetry, where $\eta$ stands for the nontrivial metric on the Hilbert space such that
\begin{eqnarray}
\hat \eta^{-1}\hat H\hat \eta = \hat H^\dagger.
\end{eqnarray}
The eigenstates with complex conjugate eigenvalues (RP resonances) must be $\eta T$-partners. Therefore, if the ground state of the model is one of the RP resonances (see Fig.\ref{Fig2}), the $\eta T$-symmetry must be spontaneously broken.

The eigenstates of the Fokker-Planck operator constitute a complete bi-orthogonal basis in the Hilbert space:
\begin{eqnarray}
\hat H \psi_n(x \chi) &=& \mathcal{E}_n \psi_n(x \chi),
\\ \bar \psi_n(x \chi) \hat H  &=& \bar \psi_n(x \chi) \mathcal{E}_n , \\
\langle\langle n |k \rangle\rangle & = & \int d^D x d^D \chi \psi_k (x \chi) \bar \psi_n (x \chi) =  \delta_{nk}, \\ \sum\nolimits_n |n \rangle\rangle \langle\langle n | &=& \sum\nolimits_n \psi_n (x \chi)\bar \psi_n (x' \chi') = \delta^D(x-x')\delta^D(\chi-\chi'),
\end{eqnarray}
where we introduced the standard notations of the models with pseudo-Hermitian evolution operators for the bra's and ket's: $| n \rangle\rangle = \psi_n(x \chi), \langle\langle n| = \bar \psi_n(x \chi) $. Bra's and ket's are related through the non-trivial metric on the Hilbert space: $\langle\langle n| = \sum\nolimits_{k}\langle k| \eta_{kn}$, where $\langle k|$ is the conventional conjugation of a ket, $\langle k| = \star \psi_k^*$, \emph{i.e.}, the combination of Hodge and complex conjugations, and the Hilbert-space metric is $(\eta^{-1})_{nk} = \langle k|n\rangle\equiv \int_X \psi_n\wedge (\star \psi_k^*)$.

Eigenstates can be either $\mathcal Q$-symmetric or not. By definition, $\mathcal Q$-symmetric eigenstates, that we call $\theta$'s, are such that $\langle\langle\theta|[\hat d,\hat X] |\theta\rangle\rangle=0$ for any $\hat X$. This requirement is equivalent to the following:
\begin{eqnarray}
\hat d |\theta \rangle\rangle = 0, \langle\langle \theta| \hat d = 0. \label{QSymmCond}
\end{eqnarray}
Clearly, all $\mathcal Q$-symmetric states have zero eigenvalues because the Fokker-Planck operator is $d$-exact.

A non-$\mathcal Q$-symmetric state does not satisfy at least one of the conditions (\ref{QSymmCond}). If, for example, $\hat H |\vartheta\rangle\rangle = \mathcal{E}_\vartheta|\vartheta\rangle\rangle$ and $|\vartheta'\rangle\rangle = \hat d|\vartheta\rangle\rangle\ne0$, then $\hat H |\vartheta'\rangle\rangle = \mathcal{E}_\vartheta|\vartheta'\rangle\rangle$ because $\hat d$ is commutative with the Fokker-Planck operator. Furthermore, $\hat d|\vartheta'\rangle\rangle=\hat d^2|\vartheta\rangle\rangle\equiv 0$ because $\hat d$ is nilpotent. The same reasoning applies to the opposite situation when $\langle\langle\vartheta|\hat H = \langle\langle\vartheta|E_\vartheta$ and $\langle\langle\vartheta|\hat d \ne 0$. In this manner, all non-$\mathcal Q$-symmetric states come in the boson-fermion (B-F) pairs, \emph{i.e.}, the pairs of states with odd and even number of ghosts.

It is also easy to see that all eigenstates with non-zero eigenvalues are non-$\mathcal Q$-symmetric. Consider an eigenstate $\hat H |\vartheta\rangle\rangle = \mathcal{E}_\vartheta|\vartheta\rangle\rangle, \mathcal{E}_\vartheta\ne 0$. If $\hat d|\vartheta\rangle\rangle\ne0$, then the pairing is obvious due to Eq.(\ref{CommutationHd}). If, in contrary, $\hat d|\vartheta\rangle\rangle=0$, when it follows immediately that $|\vartheta\rangle\rangle = \hat d |\vartheta'\rangle\rangle$, where $|\vartheta'\rangle\rangle = \hat j|\vartheta\rangle\rangle/\mathcal{E}_\vartheta$ and we used Eq.(\ref{HamiltonianTFT}). Thus, in this situation $|\vartheta\rangle\rangle$ is a member of a B-F pair too.

Each B-F pair can be parametrized by a single bra-ket pair, $\langle\langle \tilde\vartheta|$ and $|\tilde\vartheta \rangle\rangle$, in the following manner:
\begin{eqnarray}
\langle\langle \tilde\vartheta|\hat d  &,&        |\tilde\vartheta \rangle\rangle,\nonumber \\
\langle\langle \tilde\vartheta|        &,& \hat d |\tilde\vartheta \rangle\rangle,\label{nonQpairs}
\end{eqnarray}
with $\langle\langle \tilde\vartheta|\hat d|\tilde\vartheta \rangle\rangle=1$, whereas $\langle\langle \tilde\vartheta|\tilde\vartheta \rangle\rangle=0$.

The operator representation of the Witten index Eq.(\ref{TFT}) is:
\begin{eqnarray}
W = {Tr}(-1)^{\hat F}e^{-t\hat H},\label{WittenIndex:Schrodinger}
\end{eqnarray}
where
\begin{eqnarray}
\hat F = \chi^i\frac\partial{\partial\chi^i},\label{ghostnumber}
\end{eqnarray}
is the ghost number operator commutative with $\hat H$ so that it is a good quantum number:
\begin{eqnarray}
\hat F |n \rangle\rangle = F_n|n \rangle\rangle. \label{ghostnumberquantumnumber}
\end{eqnarray}
The inclusion $(-1)^{\hat F}$ appears in Eq.(\ref{WittenIndex:Schrodinger}) due to the unconventional periodic boundary conditions for the anticommuting ghosts.

The B-F pairs of the non-$\mathcal{Q}$-symmetric eigenstates from (\ref{nonQpairs}) do not contribute into the Witten index
\begin{eqnarray}
W  = \sum\nolimits_{k=0}^D (-1)^k N_k,\label{CohomologyWitten}
\end{eqnarray}
where $N_k = \#\{\theta|F_\theta=k\}$ is the number of $\mathcal Q$-symmetric eigenstates with $k$ ghosts. This expression is the STS counterpart of Eq.(\ref{LefschetzHoft1}).

As is seen from Eq.(\ref{CohomologyWitten}), the Witten index is independent of time duration, $t$. Therefore, $W$ can be evaluated, for instance, in the $t\to 0$ limit. In this limit, $M_t \to {Id}_X$ in Eq.(\ref{StochW}) for any configuration of the external noise, $\xi$. For the identity map, Eq. (\ref{LefschetzHoft1}) says that $W_{cl}(\xi)$ equals the Euler characteristic of $X$ (for closed $X$). Stochastic averaging in Eq.(\ref{StochW}) of a constant yields the same constant and we have arrived at the conclusion that the Witten index equals the Euler characteristic of the closed phase space for all smooth enough flow vector fields and for Gaussian white noises of arbitrary temperature and metric.

For more general situation of non-compact phase spaces, the Witten index must be interpreted as the winding number of the Nicolai map provided by the SDE. \cite{Mine3} In this case, it may depend not only on the topology of the phase space but also on the behavior of the flow vector field at infinities and on the choice of the class of wavefunctions that constitutes the Hilbert space.

As to the dynamic partition function, its operator representation expression has the following form:
\begin{eqnarray}
Z = {Tr} e^{-t\hat H}.\label{PartFunc:Schrodinger}
\end{eqnarray}
As compared to the Witten index, Eq.(\ref{PartFunc:Schrodinger}) is missing the topological factor $(-1)^{\hat F}$ as a result of different (antiperiodic) boundary conditions for the ghosts in Eq.(\ref{TFTZ}).

We would like to stress here that in the literature on, \emph{e.g.}, N=2 supersymmetric quantum mechanics and sigma models, it is often said that the dynamic partition function (\ref{PartFunc:Schrodinger}) is the result of the Wick rotation of the real time, $t'=i\times t$. This is not true from the point of view of stochastic dynamics. The time, $t$, is the original time of the Fokker-Planck evolution and not the Schr\"odinger evolution. This explains the absence of the imaginary unity in the exponent. The direct quantum analogue of Eq. (\ref{PartFunc:Schrodinger}) is the generating functional, $Z=Tr e^{-i t \hat H'}$, where $\hat H'$ is the Hamiltonian of the quantum model under consideration.

Another comment applies to the seeming similarity of Eq.(\ref{PartFunc:Schrodinger}) with the statistical of thermodynamic partition function of a quantum system, $Z' = Tr e^{-\hat H'/\Theta}$. The relation of $Z'$ to the story of stochastic quantization is as follows. The statistical analogue of $Z'$ is the partition function of the steady-state total probability distribution in the phase space. This probability distribution can be called the thermodynamic equilibrium (TE) state. In Sec.(\ref{Discussion}), we will discuss that for physically meaningful models such TE state always exist. Accordingly, the statistical analogue of $Z'$ is the normalization constant of the TE state.

\section{STS of GTO}
\label{GTO}
In the DS theory, there is a fundamental object of interest known as generalized TO (GTO). \cite{Ruelle2} The GTO is a way to account for the effect of stochastic noise. In this section, it is demonstrated that the STS provides a systematic framework for the formalism of the GTO, which turns out to be nothing more than the finite-time Fokker-Planck evolution operator.

\subsection{Hilbert space and exterior algebra}
Following Ref.\cite{WittenSusy}, one can identify the ghost operators with the exterior and interior multiplications:
\begin{eqnarray}
\chi^i = dx^i\wedge , \partial/\partial\chi^i= \imath_{\partial/\partial x^i}.
\end{eqnarray}
Accordingly, the Hilbert space is the (complex-valued) exterior algebra, $\mathcal H = \Omega(X)$, and wavefunctions with k ghosts are the differential k-forms on $X$, $\psi^{(k)}\in \mathcal H^k \equiv \Omega^k(X)$, so that Eq.(\ref{wavefunctionkth}) should read:
\begin{eqnarray}
\psi^{(k)}(x) = \psi_{i_1...i_k}(x) \wedge_{l=1}^{k} dx^{i_l}.\label{wavefunctionTFT}
\end{eqnarray}
In this picture, $\hat d = dx^i\wedge\partial/\partial x^i$ is the exterior derivative or the De Rahm operator, and the Fokker-Planck operator from Eq.(\ref{HamiltonianTFT}) is:
\begin{eqnarray}
\hat H = \hat {\mathcal{L}}_{F}- T \triangle,\label{HamiltonianDS}
\end{eqnarray}
where $\hat {\mathcal{L}}_{ F} = [\hat d, \imath_{ F}]_+$ is the Lie derivative representing the physical flow along the flow vector field and the operator of the noise-induced diffusion, $- \triangle = [\hat d, \hat {\bar d}],$ is a member of the family of Laplace operators. Surprisingly enough, the Laplacian of the stochastic quantization is not always the Hodge Laplacian (see Appendix).

The bra-ket combination of any eigenstate has the meaning of the total probability distribution:
\begin{eqnarray}
\psi_n \wedge \bar\psi_n = P_n(x)dx^1\wedge ...\wedge dx^D \in \Omega^D(X).
\end{eqnarray}
Bra's and ket's themselves have the meaning of marginal and conditional probability distributions. For each bra-ket pair, the marginal probability distribution is the one which is $\hat d$-closed, \emph{i.e.}, satisfies $\hat d\psi_n=0$ or $\bar\psi_n\hat d = 0$. A $\hat d$-closed differential form has no coordinate dependence in those variables, in which it is not a distribution (has no differential/ghost), just like a marginal conditional probability should. Note, that for each eigenstate, $\mathcal Q$-symmetric or not, either bra or ket satisfies this condition as is seen from Eqs.(\ref{QSymmCond}) and (\ref{nonQpairs}) (for the non-$\mathcal Q$-symmetric B-F pairs this follows also from $\hat d^2=0$). At this, for $\mathcal Q$-symmetric eigenstates both bra and ket are $\hat d$-closed. This is the situation of statistical independence. Eqs.(\ref{QSymmCond}) are necessary conditions for the possibility of the introduction of global coordinates on $X$ such that the bra and ket have no dependence on coordinates in which they are not distributions.

Let us recall now that in the deterministic limit, the bra's and ket's of $\mathcal Q$-symmetric eigenstates are the Poincar\'e duals of the global stable and unstable manifolds (up to the factors from the cohomology of the invariant manifold, on which these stable and unstable manifolds intersect). In other words, the bra/ket is a delta-functional conditional probability distribution of the unstable/stable variables transverse to the stable/unstable manifolds. \cite{Mine3} That the bra/ket of a $\mathcal Q$-symmetric eigenstate represents the unstable/stable variables must also be true for stochastic cases. From this and from the discussion in the previous paragraph, it follows that one way to understand the phenomenon of the spontaneously broken topological supersymmetry is that the stable and unstable variables are not statistically independent within the ground state.

\subsection{Definition of GTO and the inverse map}
\label{GTOPathintegral}
The GTO, $\hat{\mathcal{M}}:\mathcal{H}\to\mathcal{H}$, is defined as a stochastically averaged pullback induced by a stochastic map. In our case of a stochastic flow,
\begin{eqnarray}
\hat{\mathcal{M}}_{t} = \langle M_{-t}^*(\xi) \rangle_{Ns},\label{DefGTO}
\end{eqnarray}
where $M_{-t}^*(\xi)$ is the pullback induced by $M_{-t}(\xi)$ defined by the time-dependent flow vector field in Eq.(\ref{Fmodified}), and the notation for the stochastic averaging is introduced in Eq.(\ref{StochW}).

Let us clarify at this point why in formulas for traces the inverse map shows up, $M_{-t}$, and not the forward map $M_{t}$. The point is that the changes in wavefunctions are solely due to the coordinate transformation provided by the flow. In order to get the expression for the final wavefunction, one has to take the expression for the initial wavefunction in coordinates, $x_{init}$, and make the formal coordinate transformation to the "final" coordinates, $x_{fin}=M_{-t}(x_{init})$. This transformation is the pullback by the inverse map, $M_{-t}$.

\subsection{Unconditional existence of Q-symmetry and discrete-time dynamics}
\label{StepLike}
A pullback is a linear operator on $\mathcal H$ so that the stochastic averaging, which is essentially a weighted summation of pullbacks at different $\xi$'s, is well defined. Now, it immediately follows that the GTO is commutative with the exterior derivative because the exterior derivative commutes with any pullback:
\begin{eqnarray}
[\hat{\mathcal{M}}_{t},\hat d]_-=0.
\end{eqnarray}
The commutativity of the GTO with the exterior derivative does not depend on any assumption about the model except for the invertibility of the map (at fixed noise), which is always the case for the continuous-time dynamics. Thus, discrete-time DSs with invertible maps must also possess $\mathcal Q$-symmetry.

Only discrete-time DSs with non-invertible maps may have $\mathcal Q$-symmetry broken explicitly, \emph{i.e.}, the exterior derivative may not commute with the GTO. This explicit breakdown may turn out to be one of the ways to understand why discrete-time DSs may exhibit chaotic behavior in lower-than-three dimensional phase-spaces, unlike the continuous-time (deterministic) DSs that can be chaotic only if the dimensionality of the phase space is three and higher.

\subsection{Finite-time evolution operator}
In the deterministic limit, the equivalence between the pullback, $M_{-t}^*$, and the finite-time Fokker-Planck evolution operator is seen from the definition of the Lie derivative
\begin{eqnarray}
\partial_t M_{-t}^*(\xi)\psi(0) = - \hat{\mathcal{L}}_{F(\xi(t))} M_{-t}^*(\xi)\psi(0),\label{LieDef}
\end{eqnarray}
where the noise modified time-dependent flow vector field is from Eq.(\ref{Fmodified}) and $\psi(0)$ is any initial wavefunction. Formal integration gives
\begin{eqnarray}
M_{-t}^*(\xi) = :e^{-\int_0^t \hat{\mathcal{L}}_{F(\xi(t'))}dt'}:,\label{DetEvolution}
\end{eqnarray}
with columns denoting chronological ordering. It is now clear that Eq.(\ref{DetEvolution}) is the finite time deterministic evolution operator:
\begin{eqnarray}
\psi(t) = :e^{-\int_0^t \hat{\mathcal{L}}_{F(\xi(t'))}dt'}:\psi(0),
\end{eqnarray}
while Eq.(\ref{LieDef}) is the deterministic Fokker-Planck equation (\ref{FokkerPlanck}):
\begin{eqnarray}
\partial_t \psi(t) = - \hat{\mathcal{L}}_{F(\xi(t))} \psi(t),\label{DeterministicFP}
\end{eqnarray}
with the deterministic evolution operator obtained from Eq.(\ref{HamiltonianDS}) by setting $\Theta=0$ and by letting the flow vector field to be time-dependent.

To see that this also holds for the GTO, \emph{i.e.}, after the stochastic averaging, let us turn back to the pathintegrals' language. We use again the ghost representation of the wavefunctions in Eq.(\ref{wavefunctionkth}). Now, the deterministic pullback (with a fixed noise configuration) can be given the following operator form
\begin{eqnarray}
&M_{-t}^*(\xi,x(t),\chi(t)|x(0),\chi(0)) = \delta^D(x(0) - M_{-t}(x(t)))\nonumber \\&\times \delta^D(\chi(0) - TM_{-t}(x(0))\chi(t)),\label{PullBack}
\end{eqnarray}
so that the finite time evolution is
\begin{eqnarray}
&\psi(t, x(t),\chi(t)) = \int d^Dx(0)d^D\chi(0) M_{-t}^*(\xi,x(t),\chi(t)|x(0),\chi(0)) \nonumber \\&\times \psi(0,x(0),\chi(0)). \label{PullBackOper}
\end{eqnarray}
This expression emphasizes once again the linearity of the pullback on $\mathcal H$. It is also seen that Eq.(\ref{PullBack}) is independent of the coordinates - on invertible smooth orientation preserving transformation of coordinates the Jacobians coming from the bosonic and fermionic $\delta$-functions will cancel each other. The coordinate independence is the essence of the supersymmetric description of SDEs.

Now, one recalls the time slices picture of Sec. \ref{TFTFromTO}, (see Fig.\ref{Fig1}), introduces the Lagrange multiplier, $B$, and the antighost field, $\bar\chi$, and arrives at
\begin{eqnarray}
&M_{-t}^*(\xi,x(t),\chi(t)|x(0),\chi(0)) = \int e^{S_{cl}(\Phi,\xi)}D\Phi.
\end{eqnarray}
Here the deterministic action is from Eq.(\ref{TFTXi}), and the pathintegral is over paths that connect the "in", $x(0), \chi(0)$, and "out", $x(t),\chi(t)$, arguments of the evolution operator in Eq.(\ref{PullBackOper}).

The next step is to integrate out $\xi$'s in the same manner as it was done in Sec.\ref{SecStoch}. This will substitute $S_{cl}(\Phi,\xi)$ by the action $S(\Phi)$ defined in Eq.(\ref{newaction}):
\begin{eqnarray}
\left\langle \int e^{S_{cl}(\Phi,\xi)}D\Phi \right\rangle_{Ns} = \int e^{S(\Phi)}D\Phi.
\end{eqnarray}
At last, one can get back to the operator representation by integrating out $B$'s and $\bar\chi$'s and arrive for Gaussian white noise at
\begin{eqnarray}
\int e^{S(\Phi)}D\Phi = e^{-t\hat H}.\label{equivalence1}
\end{eqnarray}
This proves that the GTO is nothing else but the the finite-time Fokker-Planck evolution operator:
\begin{eqnarray}
\hat{\mathcal{M}}_{t} = e^{-t\hat H}.
\end{eqnarray}
The operator version of this derivation, which unambiguously resolves the Ito-Stratonovich dilemma, is given at the end of the Appendix.

\subsection{Flat traces of GTO}

The fundamental objects of study in the DS theory are the so called sharp traces (and determinants) of the GTO. Those, in turn, are defined through the so called flat traces of the GTO. The flat trace of degree $k$ is the trace of $\hat{\mathcal{M}}$ over $\mathcal{H}^k$:
\begin{eqnarray}
{Tr}^\flat_{k} \hat{\mathcal{M}}_{t}^{(k)} \equiv {Tr}_{\mathcal{H}^k} e^{-t\hat H^{(k)}} = \sum\nolimits_{F_n=k}e^{-t\mathcal{E}_n},
\end{eqnarray}
where $\hat {\mathcal{M}}_{t}^{(k)}\equiv e^{-t\hat H^{(k)}}$ together with $\hat H^{(k)}$ are projections on $\mathcal H^k$. The relation of the flat trace to the Ruelle-Frobenius-Perron TO in Sec.\ref{TOSec} can be established by considering the coordinate version of the action of the pullback on a wavefunction (\ref{wavefunctionTFT})
\begin{eqnarray}
&M_{-t}^*\psi^{(k)}(x) = \psi_{i_1...i_k}(x') \wedge_{l=1}^{k} (dx')^{i_l},\label{PullBackForms}
\end{eqnarray}
where $x' = M_{-t}(x)$ and $(dx')^{i_1} =TM_{-t}{}_{\tilde i_1}^{i_1}(x) dx^j$ is the tangent map (\ref{PushForward}) of differentials induced by $M_{-t}$ (functional dependence on the noise configuration, $\xi$, is tacitly assumed in the above formulas).

In the standard manner, the trace of Eq.(\ref{PullBackForms}) is:
\begin{eqnarray}
&{Tr}^\flat_{k} \hat{\mathcal{M}}_{t}^{(k)} = \left\langle
\sum_{x={fix} M_{-t} } \frac{{Tr} \wedge^k TM_{-t}(x)}{|det (1 - TM_{-t}(x))|}\right\rangle_{Ns}, \label{flattrace}
\end{eqnarray}
where the sum and the denominator come from the "deterministic" trace over bosonic fields, while the fermionic trace is over the extension of the tangent map (\ref{PushForward}) on the $k^{th}$ exterior power of the tangent space, $\wedge^k TM_{-t}(x): \wedge^k T_x X \to \wedge^k T_{ M_{-t}(x)} X$,
\begin{eqnarray}
{Tr} \wedge^k TM_{-t}(x)
=\sum_{i_1<...<i_k} \imath_{\partial/\partial x^{i_k}}...\imath_{\partial/\partial x^{i_1}}\wedge_{l=1}^{k} (dx')^{i_l} = m_k(x),\nonumber
\end{eqnarray}
with $m_k(x)$ from Eq.(\ref{weights:FP}). Thus
\begin{eqnarray}
{Tr}^\flat_{k} \hat{\mathcal{M}}_{t}^{(k)} = \langle{Tr}\mathcal{M}_t m_k
\rangle_{Ns}.
\end{eqnarray}

\subsection{Traces and determinants of GTO}

Finally, the so called sharp and counting (ordinary) traces of the GTO are recognized as the Witten index and the dynamic partition function:
\begin{eqnarray}
{Tr}^\sharp \hat {\mathcal{M}}_{t} &=& \sum\nolimits_{k=1}^D (-1)^k {Tr}^\flat_{k}\mathcal{M}^{(k)}_{t} = W,\label{DSVersionOFW}\\
{Tr}^c \hat {\mathcal{M}}_{t} &=& \sum\nolimits_{k=1}^D {Tr}^\flat_{k}\mathcal{M}^{(k)}_{t} = Z.
\end{eqnarray}
Complementary to these, the DS theory also deals with the sharp and counting determinants of the GTO:
\begin{eqnarray}
{Det}{}^\sharp(\hat 1 - z \hat {\mathcal{M}}_{t})^{-1}&=& \prod\nolimits_{k=0}^D {Det}{}^\flat_k(\hat 1 - z {\mathcal{M}}_{t}^{(k)})^{(-1)^{k+1}}\nonumber\\&=& \prod\nolimits_{n} (1 - z e^{ -t\mathcal{E}_n})^{ (-1)^{F_n+1}}, \label{sharpzeta} \\
{Det}{}^c(\hat 1 - z \hat {\mathcal{M}}_{t})^{-1} &=& \prod\nolimits_{k=0}^D {Det}{}^\flat_k(\hat 1 - z {\mathcal{M}}_{t}^{(k)})^{-1}\nonumber\\&=& \prod\nolimits_{n}(1 - z e^{-t\mathcal{E}_n})^{-1}.\label{zeta}
\end{eqnarray}
The notation for the determinant here is capitalized to emphasize that this is a determinant of an operator on the infinite-dimensional Hilbert space, unlike the finite-dimensional phase-space determinants we encountered previously.

\section{Discussion}
\label{Discussion}
In the previous section it was established that the operator representation of the STS is merely the GTO formalism of the DS theory. As compared to the DS theory, however, the STS provides one very important piece of understanding. This is the understanding that all the SDEs possess topological supersymmetry.

An immediate consequence of the topological supersymmetry of SDEs is the trivialization of Eqs.(\ref{DSVersionOFW}) and (\ref{sharpzeta}). Because of the B-F pairing of the non-$\mathcal Q$-symmetric states, the sharp trace of the GTO is a topological invariant, $W$. By the same token, the sharp determinant of the GTO simplifies as:
\begin{eqnarray}
{Det}{}^\sharp(\hat 1 - z \hat {\mathcal{M}}_{t})^{-1} = (1-z)^{-W}.
\end{eqnarray}

\subsection{Spectrum of the Fokker-Planck Operator}
Another utility of the existence of $\mathcal Q$-symmetry is the first theoretical explanation of the emergence of ubiquitous long-range correlations (1/f noise) in chaotic DSs. Those can be attributed to the spontaneous breakdown of $\mathcal Q$-symmetry. \cite{Mine1,Mine3} This explanation together with the picture of the $\mathcal Q$-symmetry breaking is of ultimate importance for applications. Related to the $\mathcal Q$-symmetry breaking, in turn, is the form of the spectrum of $\hat H$. Indeed, the $\mathcal Q$-symmetry is definitely broken when the ground state has non-zero eigenvalue, while which of the eigenstates are the ground states of the model is uniquely determined by the spectrum of $\hat H$ (see Fig.\ref{Fig2}). On the other hand, the DS theory provides numerous theorems that address the spectrum of GTO. Thus, the STS of the GTO established in Sec.\ref{GTO} may prove useful by shedding some additional light on the spectrum of $\hat H$ and thus on the issue of $\mathcal Q$-symmetry breaking.

One of the theorems from the DS theory \cite{Ruelle2,Ruelle1} assures that under certain conditions $\det{}^\flat _k(1 - z\hat{\mathcal M}_t^{(k)})$ for any $k$ has no poles (is meromorphic) for $|z| < e^{-P}$, where $P$ is some model specific constant related to a parameter called pressure.

As is seen from Eq.(\ref{zeta}), the logarithms of the positions of the poles of the GTO's determinants correspond to the eigenvalues of $\hat H$. Therefore, the theorem seemingly assures that the attenuation rates of the eigenvalues in Eq.(\ref{Eigenvalues}) are bounded from below. \footnote{From various theorems of the classical DSs theory is follows that this is true even in the deterministic limit, where the Laplacian vanishes and the Fokker-Planck operator looses its ellipticity.}\footnote{That the real part of eigenvalues are bounded from below must be always true for models with  the noise-metric which is positive definite everywhere because the Fokker-Planck operators for such models is elliptic.} In other words, for the ground state $\Gamma_g = \min_n \Gamma_n > -\infty$. In many cases spectral theorems also suggest that at $|z| = e^{-P}$ there is a single real pole at $z = e^{-P}$. This statement has been supported by numerical analysis of some DSs (see, \emph{e.g.}, Ref.\cite{RPResonances}). In terms of the spectrum of the Fokker-Planck operator this means that the ground states eigenvalue is real, $\mathcal{E}_g=\Gamma_n$. The same picture seems to appear from the physical arguments in the forthcoming discussion.

\subsubsection{Unbroken $\mathcal Q$-symmetry: thermodynamic equilibrium}
\label{TEstate}
One of the requirement on $\Gamma_g$ is that it can not be positive. This is straightforwardly seen for models with non-zero Witten index, which must possess $\mathcal Q$-symmetric state(s) of zero eigenvalue. In fact, one zero-eigenvalue $\hat d$-symmetric eigenstate from $\Omega^D(X)$ must always exist for physically meaningful models. To see this, let us note that all non-$\hat d$-symmetric eigenstates from $\Omega^D(X)$ are in fact $\hat d$-exact. Indeed, such states come in the B-F pairs of the form, $|\tilde\vartheta\rangle\rangle$ and $\hat d|\tilde\vartheta\rangle\rangle$ (see Eq.(\ref{nonQpairs})). Therefore, all members of the B-F pairs from $\Omega^D(X)$ can only be of the form $\hat d|\tilde\vartheta\rangle\rangle$. The integral of such wavefunctions over the phase space is zero: $\int_X\hat d|\tilde\vartheta\rangle\rangle=\int_{\partial X}|\tilde\vartheta\rangle\rangle = 0$, provided that $\partial X=0$ or that the wavefunction is zero at the boundary of the phase space or at the spatial infinity for non-compact phase spaces.

On the other hand, any physical wavefunction from $\Omega^D(X)$ must be such that $\int_{X}\psi=1$. The meaning of this requirement is that the probability of finding the DSs in the entire phase space must be unity. Clearly, any physical wavefunction can not be resolved in the eigenstates of $\hat H$ unless there is at least one eigenstate, whose integral over the phase space is non-zero. As follows from the discussion in the previous paragraph, such eigenstate must be $\mathcal Q$-symmetric and thus be of zero-eigenvalue.

This stationary (zero-eigenvalue) eigenstate is referred sometimes to as "ergodic zero". We prefer to call it the state of thermodynamic equilibrium (TE), \emph{i.e.}, the state when the DS has forgotten its initial conditions and is represented by a stationary total probability distribution. If this eigenstate is (one of) the ground state(s) as in Fig.(2a), the $\mathcal Q$-symmetry is not broken spontaneously and the DS can be said to be at its TE.

The unconditional existence of the TE state may turn out to be a part of a more general statement that (for closed phase spaces) each De Rahm cohomology class provides one $\mathcal Q$-invariant eigenstate. This statement can be proven for example for models, in which the diffusion Laplacian is the Hodge Laplacian (see the discussion in Appendix), and in the large temperature limit, $\Theta\to\infty$, where one can utilize the conventional perturbation theory. In this limit, the zeroth order Fokker-Planck operator is the Hodge Laplacian, the $\mathcal Q$-invariant eigenstates are the harmonic differential forms from the De Rahm cohomology, and all the other states are non-$\mathcal Q$-invariant and have positive and real eigenvalues. The Lie derivative along the flow vector field is a perturbation of order $\Theta^{-1}$. It is straightforward to demonstrate that to all orders in $\Theta^{-1}$, all the harmonic forms, $|\theta_{dR}\rangle\rangle$, acquire $\hat d$-exact corrections: $|\theta_{dR}\rangle\rangle \to |\theta_{dR}\rangle\rangle + \hat d |{something}\rangle\rangle$, so that they are still $\mathcal Q$-invariant eigenstates of zero eigenvalue.

There is also a possibility that both $\mathcal Q$-symmetric and non-$\mathcal Q$-symmetric ground states exists simultaneously. This can happen when some of the B-F pairs of states accidently have zero eigenvalue. In this situation one may appeal to the standard argument that different ground states correspond to different physical realizations of the model. Within this picture, vacuum expectation values in Eq.(\ref{VEV}) below must be reduced to the contributions from only that "realized" ground state. Having mentioned this accidental possibility, from now on we believe that if the $\mathcal Q$-symmetry is not spontaneously broken then non-$\mathcal Q$-symmetric states are not among the ground states of the model.

The TE state has the following general form, $e^{-L/T}\star 1$, where $\star 1$ is the invariant volume on $X$ and function $L$ can be recognized as the stochastic generalization of the Lyapunov function from the theory of deterministic DSs. For Langevin SDEs, the role of the generalized Lyapunov function is played by the Langevin potential. \cite{Mine2} In the literature on N=2 supersymmetric quantum mechanics (SQM), one may encounter examples of superpotentials (or Langevin functions) that provide non-integrable TE states. The simplest example of such models is the N=2 SQM on $\mathbb{R}^1$ with the superpotential going to minus infinity at both spatial infinities. Such a model makes sense only from the point of view of metastable dynamics, that is, on the level of the perturbative ground states around local minima of the superpotential. From the point of view of the global ground state of the model and/or of the long-time dynamics, this model is not physical because the DS will eventually escape to (one of) the spatial infinity(ies) where the superpotential goes to $-\infty$ and will never come back. In fact, the existence of the integrable TE state can be viewed as a condition for the model to be physical.

\subsubsection{Spontaneously broken $\mathcal Q$ symmetry: stochastic chaos}
\label{SecChaos}

In Ref.\cite{Mine3}, the $\mathcal Q$-symmetry breaking picture was discussed under the assumption that $\Gamma_g$ can not be negative (see Fig.\ref{Fig2}b). This assumption was based on the following argument. The Witten index can be viewed as the partition function of the noise, and since the noise does not have instabilities, states with negative attenuation rates must not exist. This argument, however, must be discarded because the eigenstates with nonzero eigenvalues are non-$\mathcal Q$-symmetric and consequently they do not contribute to the Witten index representing (up to a topological factor) the partition function of the noise.

This does not necessarily suggest, however, that reasons forcing $\Gamma_g$ to be nonnegative do not exist in some classes of DSs. In DSs that do have such a reason, Fig.\ref{Fig2}b is the only possible picture of the $\mathcal Q$-symmetry breaking. In general case, on the other hand, $\Gamma_g$ can be negative. The three possible spectra corresponding to this situation are given in Figs.\ref{Fig2}c,\ref{Fig2}d, and \ref{Fig2}f.

For the situation in Fig.\ref{Fig2}d, the partition function can take on negative values in the long time limit, $Z|_{t\to\infty} \sim e^{t |\Gamma_g|}\cos E_g t$. The same is true for Fig.\ref{Fig2}d. If we recall now that in the long time limit $Z$ must have the meaning of averaged number of periodic solutions/orbits (see Secs. \ref{PhysPartFunc} and \ref{TOSec}), the negativeness of $Z$ looks suspicious. This, however, does not necessarily point onto the possibility that such FP spectra are not realizable. It only suggests that for such models the dynamic partition function does not represent the number of periodic solutions. 

In either case, Fig.\ref{Fig2}c seems to be most likely picture of $\mathcal Q$-symmetry breaking. In this situation, \footnote{Factor 2 here comes from the two-fold degeneracy of the non-$\mathcal Q$-symmetric BF pair of the ground state}
\begin{eqnarray}
\lim_{t\to\infty} Z \approx 2 e^{t |\Gamma_g|}.\label{TopologicalEntropy}
\end{eqnarray}
In other words, the stochastically averaged number of periodic solutions/orbits grows exponentially in the large time limit with rate $|\Gamma_g|$. \footnote{Parameter $|\Gamma_g|$ can be identified as the previously mentioned pressure of the DS. Yet another parameter to which $\Gamma_g$ can be related is the concept of (topological) entropy of a chaotic DS.} This exponential growth is a unique feature of deterministic chaos and Eq.(\ref{TopologicalEntropy}) is the stochastic generalization of this situation (in the deterministic limit, this growth is provided by the infinite number of unstable periodic orbits with arbitrary large periods constituting strange attractors). The picture we just arrived at proves that the spontaneous breakdown of the topological supersymmetry is indeed the field-theoretic essence of the concept of deterministic chaos that also provides the stochastic generalization for this concept.

The ground state of a chaotic DSs is not a total probability distribution. It is not a distribution in the unstable/unthermalized variables, in which the DS has infinitely long memory of initial condition/perturbations. In the theory of deterministic dynamics, these unstable variables are featured by positive Lyapunov exponents. On the other hand, just like in quantum mechanics, it is the bra-ket combination (of the ground state) which is the total probability distribution. In a sense, the dynamics factorizes the total probability distribution into two differential forms that are the conditional/marginal probability densities for the stable (ket) and unstable (bra) variables.

\subsection{Response and the butterfly effect}
\label{SecButterflyEffect}
The physical way to couple the DS to external influence is to introduce of a set of probing fields, $\phi^c(t)$, into the flow vector field:
\begin{eqnarray}
F^i(x(t)) \to F^i(x(t)) + \phi^c(t) f^i_c(x(t)),
\end{eqnarray}
where $f$'s is a set of some vector fields on $X$. The action transforms as:
\begin{eqnarray}
S \to S + \int dt \phi^c(t) \left\{ {\mathcal Q}, i\bar\psi_i(t) f^i_c(x(t))\right\}.
\end{eqnarray}
The response of the DS to the perturbations can now be characterized by the set of the following stochastic expectation values: (here we consider only spectra of type \ref{Fig2}a and c)
\begin{eqnarray}
&& \overline{\prod\nolimits_{k=1}^l \left\{ {\mathcal Q}, i\bar\psi_{k_1}(t_1) f^{k_1}_{c_k}(x(t_k))\right\}}=\left.Z^{-1}\left(\prod\nolimits_{k=1}^l\frac{\delta}{\delta \phi^{c_k}(t_k)}\right)Z\right|_{\phi=0}\nonumber\\&&
=Z^{-1}\int_{APBC} e^{S}\prod\nolimits_{k=1}^l \left\{ {\mathcal Q}, i\bar\psi_{i_k}(t_k) f^{i_k}_{c_k}(x(t_k))\right\}\nonumber\\&& =Z^{-1}\sum\nolimits_n \int\langle\langle n| e^{S}\prod\nolimits_{k=1}^l \left\{ {\mathcal Q}, i\bar\psi_{i_k}(t_k) f^{i_k}_{c_k}(x(t_k))\right\}|n\rangle\rangle.
\end{eqnarray}
The summation in the last line is over all the eigenstates, $n$, and the pathintegration connects the arguments of bra's and ket's.

Of primary importance is the limit, $t\to \infty$, where the stochastic expectation values become the vacuum expectation values (VEV's):
\begin{eqnarray}
&& \left.\overline{\prod\nolimits_{k=1}^l \left\{ {\mathcal Q}, i\bar\psi_{k_1}(t_1) f^{k_1}_{c_k}(x(t_k))\right\}}\right|_{t\to\infty}\nonumber\\&& =N_g^{-1}e^{t\mathcal E_g}\sum\nolimits_g \langle\langle g| e^{S}\prod\nolimits_{k=1}^l \left\{ {\mathcal Q}, i\bar\psi_{i_k}(t_k) f^{i_k}_{c_k}(x(t_k))\right\}|g\rangle\rangle.\label{VEV}
\end{eqnarray}
The summation here is over the ground states only and we used that $Z|_{t\to\infty} = N_ge^{-t\mathcal E_g}$ with $N_g$ being the number of ground states and $\mathcal E_g$ being the ground states' eigenvalue. The exponential factor compensates for the same factor that comes from the time propagation of the ground states.

Clearly, all the VEV's of $\mathcal Q$-exact operators vanish when the $\mathcal Q$-symmetry is not broken spontaneously and all the ground states are $\mathcal Q$-symmetric. This can be interpreted as though the DS forgets all the perturbations in the long-time limit. In the opposite situation of spontaneously broken $\mathcal Q$-symmetry, some (or rather most of) VEV's of $\mathcal Q$-exact operators do not vanish. Accordingly, this situation can be interpreted as though the chaotic DS does not forget the perturbations even in the limit of infinitely long temporal evolution. In this way, the STS reveals the famous butterfly effect or the chaotic sensitivity to initial conditions/perturbations.

It would not be surprising if it turned out that some of the VEV's of $\mathcal Q$-exact operators are of topological origin. If this is indeed true, the topological nature of such VEV's must be conceptually different from the conventional topological invariants of BPS observables on instantons (see the discussion in Sec.\ref{BeyondInstantons}).

\begin{figure}[th]
\begin{center}
\includegraphics[width=10cm,height=5cm]{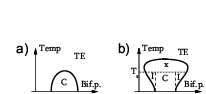}
\end{center}
\caption{\label{Fig3} Two qualitatively different phase diagrams. The axes are the intensity (or temperature) of the noise and the other "bifurcation" parameters. {\bf (a)} In this simple type of DSs, there are only two phases: the thermodynamic equilibrium phase (TE) with unbroken $\mathcal Q$-symmetry, and the ordinary chaotic phase (C) with $\mathcal Q$-symmetry spontaneously broken in the low temperature limit by the non-integrability of the flow vector field. The only effect of the noise is to restore the $\mathcal Q$-supersymmetry at high temperatures.  {\bf (b)} In this second type of DSs, the low-temperature limit there appears the second phase that can be called the noise-induced (or intermittent) chaos (I) with the $Q$-symmetry spontaneously broken by the condensation of (anti-)instantons. Above certain temperature, $T_{x}$, the sharp boundary between the I- and C- phases must get smeared into a crossover and the two phases must merge into a phase indicated as $x$. At even higher temperatures the $Q$-symmetry must get restored.}
\end{figure}

\subsection{General phase diagram}
\label{GeneralPhaseDiagram}
As a finalizing remark, we would like to revisit here the issue of the generic phase diagram. This subject matter is of a somewhat speculative character. Nevertheless, it is still worth a discussion as the phase diagram makes a good sense from the physical point of view and the logic behind it looks robust.

It was proposed in Ref. \cite{Mine2} that there must exist three major phases: the TE phase with unbroken $\mathcal Q$-symmetry that was previously identified as Markovian in order to emphasize that such DSs do not exhibit long-range chaotic order; the noise-induced chaotic phase (intermittent phase) with $\mathcal Q$-symmetry broken by the condensation of the configurations of instantons and antiinstantons; and the conventional chaotic phase, where the $\mathcal Q$-symmetry is broken even in the deterministic limit due to the non-integrability of the flow vector field.

The logic behind the introduction of this phase diagram is as follows. In the deterministic limit, the phase diagram is divided into two regions of integrable deterministic dynamics and non-integrable chaotic deterministic dynamics. The effect of noise provides two additional sources of $\mathcal Q$-symmetry breaking for the DSs with integrable flow vector fields. The first source is the perturbative corrections. Supersymmetries, however, are hard to break perturbatively (non-renormalization theorems). This is actually the reason why the second source of the condensation of the configurations of (anti-)instantons is considered one of the most reliable mechanisms of the spontaneous breakdown of supersymmetries in the high-energy physics models. \cite{WittenSuSyBreakingInstantons}

In terms of stochastic dynamics, configurations of (anti-)instantons represent in particular the noise-induced tunneling processes between different deterministic attractors. Such processes exist even in Langevin SDEs with multiple local minima of Langevin potential. At this, $\mathcal Q$-symmetry is never broken for (physically meaningful) Langevin SDEs (the FP spectra are real and non-negative). The conclusion is that the very existence of (anti-)instantons does not necessitate the spontaneous breakdown of $\mathcal Q$-symmetry. In a sense, condensation of (anti-)instantons can only "help" $\mathcal Q$-symmetry breaking in DSs that are already "close" to be chaotic in the deterministic limit.

Another important observation is that in the deterministic limit anti-instantons disappear because they are the noise-induced processes of going against the flow vector field (as a result antiinstantonic matrix elements have exponentially weak factors such as Gibbs factors vanishing in the deterministic limit). At the same time, instantons on their own can not condense into the global ground state because instantons correspond to classical solutions leading from less stable invariant manifolds to more stable ones and without "compensation" from anti-instnatons they can not realize the periodic boundary conditions (see, however, the discussion of the out-of-equilibrium dynamics in quenches in Sec. 3.4). As a result, when anti-instantons disappear so does the possibility of the noise-induced $\mathcal Q$-symmetry breaking.

The only possible picture that follows from the discussion in the two previous paragraphs is this. Spontaneous $\mathcal Q$-symmetry breaking by the condensation of (anti-)instantons (noise-induced chaos) is possible only close to the transition into the ordinary deterministic chaos. In the deterministic limit, this phase collapses into the boundary of the conventional deterministic chaotic phase. For an external observer, the noise-induced chaotic behavior must look like an intermittent dynamics where the DS spends most of its time on deterministic attractors, jumping sporadically from one attractor to another. At this, these jumps must exhibit the signatures of a long-range chaotic order such as algebraic statistics.

The above reasonings work only in the low-temperature limit, where the dominant part of the Fokker-Planck operator is the drift term (the Lie derivative along the flow vector field), the perturbative ground states are localized on the unstable manifolds of the flow vector field, while the presence of the noise somewhat smears these ground states. It is only in this regime that these ground states overlap insignificantly making the concept of (anti-)instantons physically sound, \emph{i.e.}, an external observer would be able to tell one process from another.

In the high temperature limit, the diffusive (Laplacian) part of the Fokker-Planck operator becomes more and more dominant as one rises the temperature. This suggests that in those SDEs, in which the diffusion Laplacian alone does not break the $\mathcal Q$-symmetry (this is certainly true for Hodge Laplacians), the $\mathcal Q$-symmetry eventually gets restored as one rises the temperature. It can be shown more rigorously for closed phase spaces, where in order to prove that for sufficiently high $\Theta$s the $\mathcal Q$-symmetry is unbroken one can use the perturbation theory in the same manner as it was used in Sec.\ref{TEstate} to discuss the relation between the supersymmetric states and De Rahm cohomology classes.

One of the ways to physically interpret this situation is that chaotic long-range order can always be destroyed by a sufficiently strong noise. The scenario when the ground state's eigevalue tends but never reaches zero with the increase in the temperature can probably be also realizable in some classes of DSs. This scenario can be thought of a special case when the boundary of the $\mathcal Q$-broken phase moved to infinity on the temperature scale.

Another issue is the boundary between the noise-induced and ordinary chaos, which is not a topological supersymmetry breaking phase transition even in the low-temperature limit (one can not break topological supersymmetry twice). In the high-temperature limit, this boundary must smear out into a crossover because the concept of (anti-)instantons is not very solid anymore as the perturbative ground states overlap significantly. In other words, the two $\mathcal Q$-broken phases must merge into a single complicated phase with a spontaneously broken $\mathcal Q$-symmetry. We did not manage to find analogues for this phase in the literature and for this reason we simply called it the x-phase on the emerging phase diagram in Fig.\ref{Fig3}b.

We would also like to point out that the noise-induced chaotic phase may exist only in DSs with a rich "enough" instantonic structure, \emph{e.g.}, with a multitude of attractors. In cases where the instantonic structure is not rich enough, \emph{e.g.}, a chaotic DS with only one strange attractor with the attraction basin being the whole phase space, the noise-induced chaotic phase must not exist. For such DSs, the phase diagram must look like this (see. Fig.\ref{Fig3}a): the ordinary chaotic phase gradually narrows down with the increase of temperature until it completely disappears.

\section{Conclusion}\label{Conclusion}

In this paper, the connection is established between the transfer operator formalism of the dynamical systems' theory and the recently proposed supersymmetric theory of stochastic differential equations. The established connection provides a potential for a fruitful cross-fertilization between the developments in the dynamical systems theory and supersymmetric and cohomological field theories. Three distinct results enabled by this connection were presented in this paper.

First, it became possible to apply the spectral theorems of the dynamical systems theory to the spectrum of the Fokker-Planck operator and refine the previously proposed picture of the spontaneous breakdown of the topological supersymmetry. Specifically, it allowed to extend the picture onto situations when the ground state's attenuation rate is negative. In these situations, the stochastically averaged number of periodic solutions/orbits grows exponentially in the large time limit, which is a unique feature of chaotic behavior. Hence, this constitutes a firm evidence proving that the spontaneous breakdown of the topological supersymmetry is indeed the field-theoretic definition and stochastic generalization of the concept of deterministic chaos.

Second, it was shown that the so called sharp trace and sharp determinant of the generalized transfer operator of the dynamical systems theory is subject to the supersymmetric trivialization due to the unconditional existence of the topological supersymmetry.

Third, this connection enabled the utilization of the Lefschetz theorem for the proof of that the Witten index of an SDE with a closed phase space equals the Euler characteristic of the phase space. To the best of our knowledge, this is the first proof of this statement suitable for any smooth enough flow vector fields, noise temperatures, and the noise metrics.

In addition, in this paper we revisited the question of the generic phase diagram previously analyzed in the low-temperature limit, where it consists of the three major phases: thermodynamic equilibrium, noise-induced chaos ((anti-)instanton condensation, intermittent), and ordinary chaos (non-integrability of the flow vector field). This picture was generalized to the high temperature regime, where the sharp boundary between ordinary and noise-induced chaotic phases must smear out into a crossover and at even higher temperatures the topological supersymmetry must get restored. It was also demonstrated that the correct Fokker-Planck temporal evolution is provided by the Weyl-Stratonovich interpretation of the SDEs, which resolves the Ito-Stratonovich dilemma.


\section*{Acknowledgements}

We would like to thank Kang L. Wang for encouragement and discussions. We are also grateful to Torsten En{\ss}lin for valuable comments and suggestions. The work was partially supported by DARPA Contract No. HR0011-01-1-0008.

\appendix
\section*{Appendix: Stochastic Weyl-Stratonovich  quantization}
\label{WeylQuantSec}
The goal of this section is two-fold. First, it provides a detailed derivation of the Fokker-Planck operator. Second, at the end of this section the Ito-Stratonovich dilemma is discussed and resolved unambiguously in favor of the Weyl-Stratonovich approach.

The stochastically averaged infinitesimal evolution of a wavefunction between two time slices, say $0$ and $\Delta t\to0$, and for the Gaussian white noise can be expressed in the following manner:
\begin{eqnarray}
\psi(x_1\chi_1\Delta t) &=&\nonumber \left\langle \int d^D x_0d^D \chi_0 \delta^D (x_0-M_{-\Delta t}(x_1))\nonumber\right.\\ &&\times \left.
\delta^D (\chi_0-TM_{-\Delta t}\chi_1)\right\rangle_{Ns}\psi(x_0\chi_0 0),\label{Eq1Noname}
\end{eqnarray}
where the averaging over the noise variable is defined as:
\begin{eqnarray}
\langle ... \rangle_{Ns} = c \int d^D \xi_0 e^{-\Delta t (\xi_0^a)^2/2} ...
\end{eqnarray}
with $c^{-1} = \int d^D \xi_0 e^{-\Delta t (\xi_0^a)^2/2}$ being the normalization constant. Eq.(\ref{Eq1Noname}) is the stochastic average of the infinitesimal pullback from Eq.(\ref{PullBack}).

Now, one can use the following discrete-time approximation of the SDE:
\begin{eqnarray}
x_0 = x_1 - \Delta t F(x^+,\xi_0), x^+ = (x_1+x_0)/2,\label{Eq2Noname}
\end{eqnarray}
where the flow vector field is defined in Eq.(\ref{Fmodified}). Similarly, the approximate evolution for the ghosts is
\begin{eqnarray}
\chi_0 = \chi_1 - \Delta t \hat{\mathbf F} (x^+,\xi_0)\chi^+,  \chi^+ = (\chi_1+\chi_0)/2,\label{Eq2Noname1}
\end{eqnarray}
with
\begin{eqnarray}
\nonumber
\hat{\mathbf F}^i_j(x,\xi) = \partial F^i(x,\xi)/\partial x^j=\partial F^i(x)/\partial x^j + (2T)^{1/2} (\partial e^i_a(x)/\partial x^j) \xi^a(t).
\end{eqnarray}
The meaning of Eqs.(\ref{Eq2Noname}) and (\ref{Eq2Noname1}) is that the flow vector field must be evaluated at the middle of the transition time interval $(0,\Delta t)$. This is the best discrete-time approximation of a continuous-time SDE and this is the origin of the necessity to use the Weyl symmetrization below. In the literature on SDEs, Weyl quantization is known also as Stratonovich approach to SDEs and in the end of this section we will discuss it in detail.

Eqs.(\ref{Eq2Noname}) and (\ref{Eq2Noname1}) implicitly define $x_0$ and $\chi_0$ in terms of $x_1$ and $\chi_1$. The bosonic $\delta$-function in Eq.(\ref{Eq1Noname}) can be given with the help of Eq.(\ref{Eq2Noname}) as:
\begin{eqnarray}
\delta^D (x_0-M_{-\Delta t}(x_1)) = |C|^{-1} \delta^D (x_0 - (x_1 - \Delta t F(x^+,\xi_0))),
\end{eqnarray}
where $C=\det(\hat 1 + \Delta t \hat {\mathbf F}/2)\sim1$. This factor is compensated by the same factor in the nominator provided by the fermionic $\delta$-function:
\begin{eqnarray}
\delta^D (\chi_0-TM_{-t}\chi_1) = C \delta^D (\chi_0 - (\chi_1 - \Delta t \hat{\mathbf F} (x^+,\xi_0)\chi^+)).
\end{eqnarray}
Now we can introduce the Lagrange multiplier, $B_0$, and the antighost, $\bar\chi_0$, in order to exponentiate the $\delta$-functions and integrate out the noise variable:
\begin{eqnarray}
\psi(x_1\chi_1\Delta t) &=& \int d^D x_0 d^D\left(\frac{B_0}{2\pi}\right) d^D \chi_0 d^D(i\bar\chi_0) \nonumber\\ &&\times e^{iB_0(x_1-x_0) - i\bar\chi_0(\chi_1-\chi_0) 
- \Delta t H(B_0 \bar\chi_0 x^+ \chi^+)}\psi(x_0\chi_00),\label{Eq3Noname}
\end{eqnarray}
where
\begin{eqnarray}
H(B \bar\chi x \chi ) &=& (i B )_i F^i(x ) - (i\bar\chi )_i F^i_{'j}(x )\chi ^j - T u_a u_a,
\end{eqnarray}
with $u_a = (iB)_i e^i_a(x)-(i\bar\chi)_ie^i_{a'j}(x)\chi^j$, is nothing more than the Fokker-Planck Hamilton function defined by Eqs.(\ref{PathIntegralH}) and (\ref{GaugeCurrent}).

Taylor expanding Eq.(\ref{Eq3Noname}) in $\Delta t\to 0$ we get ($\Phi_0\to\Phi'$, and $x_1,\chi_1\to x,\chi$):
\begin{eqnarray}
\partial_t \psi(x\chi t) &=& \nonumber{lim}_{\Delta t\to0} (\psi(x\chi\Delta t) -\psi(x\chi0))/\Delta t\\ &=& - \nonumber\int d^D x' d^D\left(\frac{B'}{2\pi}\right)d^D\chi'd^D(i\bar\chi') e^{iB'(x-x') - i\bar\chi(\chi-\chi')} \\ &&\times H(B',\bar\chi',(x+x')/2,(\chi+\chi')/2) \psi(x'\chi' t).\label{WeylIntegral}
\end{eqnarray}
Now it is clear that the Lagrange multiplier $B'$ is the momentum operator diagonal in the Fourier space provided by the integration over $x'$, while the subsequent inverse Fourier transform (the later integration over $B'$) brings the wavefunction back again into the representation where $x$ is diagonal. The same is true for the fermionic momentum $\bar\chi$. This is the reason for the standard operator identification in Eqs.(\ref{AdditionalFields}).

In fact, there is a seeming sign ambiguity in the antighost operator: $i\hat{\bar\chi}_i = \pm \partial/\partial\chi^i$. This ambiguity can be removed by demanding that the bi-graded commutation of the exterior derivative, $\hat d$, \emph{i.e.}, the N\"other charge of the $\mathcal Q$-symmetry, acts on the fields' operators in the same way as $\mathcal Q$-operator acted on their pathintegral versions:
\begin{eqnarray}
\{{\mathcal Q}, (x^i,B_i,\chi^i,\bar\chi_i)\} = (\chi^i,0,0,B_i).
\end{eqnarray}
As can be straightforwardly verified, with the sigh chosen in Eq.(\ref{AdditionalFields}) this condition is satisfied:
\begin{eqnarray}
\left[\hat d, (\hat x^i,\hat B_i,\hat \chi^i,\hat{\bar\chi}_i) \right] = (\hat \chi^i,0,0,\hat B_i),\label{Qsym}
\end{eqnarray}
where the bi-graded commutation is defined as a commutator for bosonic operators and anti-commutator for fermionic ones.

Due to the symmetric way the coordinates $x$'s and $x'$'s ($\chi$'s and $\chi'$'s) enter the argument of the Fokker-Planck Hamilton function, on substituting momenta by their operator versions one must perform the Weyl symmetrization. Indeed, suppose the Fokker-Planck Hamilton function has a term, $- iB_iG^i(x)$, where $G^i$ is some function of $x$. This term will contribute the following expression into the r.h.s. of Eq.(\ref{WeylIntegral}) (fermionic fields are not shown):
\begin{eqnarray}
&& \int d^D x' d^D\left(\frac{B'}{2\pi}\right) e^{iB'(x-x')} iB'_iG^i((x'+x)/2)\psi(x't)\nonumber\\
&=& \frac\partial{\partial x^i} \int d^D x' d^D\left(\frac{B'}{2\pi}\right)e^{iB'(x-x')} G^i((x'+x)/2)\psi(x't)\nonumber\\
&& - \frac12 \int d^D x' d^D\left(\frac{B'}{2\pi}\right)e^{iB'(x-x')} \left(\frac\partial{\partial x^i}G^i((x'+x)/2)\right)\psi(x't)\nonumber\\
&=&\frac12\left(\frac\partial{\partial x^i} G^i(x) + G^i(x) \frac\partial{\partial x^i} \right)\psi(xt).\nonumber
\end{eqnarray}
The same is true for fermionic fields with the only difference that the symmetrization must be fermionic, \emph{e.g.}, $\left[\hat {\bar \chi}_j, \chi^i \right]_{sym} = (1/2)(\hat {\bar \chi}_j \chi^i - \chi^i \hat {\bar \chi}_j )$. In general case, the bi-graded symmetrization must be used:
\begin{eqnarray}
\left[\hat A, \hat B \right]_{sym} = (1/2)(\hat A \hat B \pm \hat B\hat A),
\end{eqnarray}
where minus sign is used when both operators are fermionic, \emph{i.e.}, have odd number of $\chi$'s and $\bar\chi$'s, and the plus sign is for the other three possibilities.

To simplify the procedure of finding the operator expression for the Fokker-Planck operator, one can make use of the "commutativity" of the operation of the bi-graded symmetrization and the substitution of the momenta fields by their operator expressions:
\begin{eqnarray}
\left[\left.X(\Phi)\right|_{B\bar\chi\to\hat B \hat{\bar\chi}}\right]_{sym} = \left.\left[X(\Phi)\right]_{sym}\right|_{B\bar\chi\to\hat B \hat{\bar\chi}},
\end{eqnarray}
where the bi-graded symmetrization of $X(\Phi)$ has the same rules described above with the only difference that the fields $\Phi$'a are not operators but the Grassmann and c-numbers. One also notices that
\begin{eqnarray}
\left[\{{\mathcal Q}, X_1 X_2...\}\right]_{sym} = \left\{{\mathcal Q}, \left[ X_1X_2...\right]_{sym}\right\},
\end{eqnarray}
and that
\begin{eqnarray}
\left. \{{\mathcal Q}, X \}\right|_{B\bar\chi\to\hat B \hat{\bar\chi}}
=\left[\hat d, \left. X \right|_{B\bar\chi\to\hat B \hat{\bar\chi}}\right],
\end{eqnarray}
as follows from Eq.(\ref{Qsym}). With the use of these properties and Eq.(\ref{PathIntegralH}) one has
\begin{eqnarray}
\hat H = \left[\left.H(\Phi)\right|_{B\bar\chi\to\hat B \hat{\bar\chi}}\right]_{sym}= \left[\hat d, \hat j\right],\label{HamiltAdditional}
\end{eqnarray}
where
\begin{eqnarray}
\hat j &=& \left[\left.j(\Phi)\right|_{B\bar\chi\to\hat B \hat{\bar\chi}}\right]_{sym},
\end{eqnarray}
with $j(\Phi)$ given in Eq.(\ref{GaugeCurrent}). One can now proceed straightforwardly:
\begin{eqnarray}
\hat j &=& \left[\frac{\partial}{\partial \chi^i} \left(F^i(x) - \Theta g^{ij}\left(\frac{\partial}{\partial x^j} -\tilde\Gamma_{kj}^l\chi^k\frac{\partial}{\partial \chi^l}\right)\right)\right]_{sym}\nonumber\\
&=&
\frac\partial{\partial \chi^i} F^i(x) - \Theta \left(\frac\partial{\partial \chi^i}\frac12\left(g^{ij}\frac\partial{\partial x^j}+\frac\partial{\partial x^j}g^{ij}\right) \right.\nonumber\\ &&\left.+ g^{ij}\tilde\Gamma_{kj}^l\frac12\left(\frac\partial{\partial \chi^i}\frac\partial{\partial \chi^l} \chi^k + \chi^k\frac\partial{\partial \chi^i}\frac\partial{\partial \chi^l}\right) \right),\nonumber\\
&=&
\frac\partial{\partial \chi^i} \left(F^i(x) - \Theta \left(g^{ij}\left(\frac\partial{\partial x^j} - \tilde\Gamma_{kj}^l\chi^k \frac\partial{\partial \chi^l} \right) + \frac12 (\tilde \nabla_{j}g^{ij})\right)\right),\label{operatorj}
\end{eqnarray}
where
\begin{eqnarray}
\tilde \nabla_{j}g^{ij} = (\partial g^{ij}/\partial x^j)+ \tilde\Gamma^l_{lj}g^{ji} + \tilde\Gamma^i_{lj}g^{jl} = 0,
\end{eqnarray}
because of the metric compatibility of the Weitzenb\"ock connection (see Eq.(\ref{MetricCompat})). Now, in agreement with the definitions introduced for Eq.(\ref{HamiltonianTFT})
\begin{eqnarray}
\hat j & = &  \frac\partial{\partial \chi^i} F^i(x) + \Theta \hat {\bar d},\label{jadditional}
\end{eqnarray}
with
\begin{eqnarray}
\hat {\bar d} = -\frac\partial{\partial \chi^i} g^{ij}\hat { \tilde \nabla}'_{j},\label{adjoint}
\end{eqnarray}
and
\begin{eqnarray}
\hat {\tilde \nabla}'_{j} = \frac\partial{\partial x^j} - \tilde\Gamma_{kj}^l\chi^k \frac\partial{\partial \chi^l}=\hat {\tilde \nabla}_{j} - T_{jk}^l\chi^k \frac\partial{\partial \chi^l},\label{CovariantDerivatTorsion}
\end{eqnarray}
where
\begin{eqnarray}
\hat {\tilde \nabla}_{j} &=& \frac\partial{\partial x^j} - \tilde\Gamma_{jk}^l\chi^k \frac\partial{\partial \chi^l},\\
T^{l}_{jk}&=& \tilde\Gamma_{kj}^l - \tilde\Gamma_{jk}^l,
\end{eqnarray}
are respectively the covariant derivative operator and the torsion of the Weitzenb\"ock connection.

Operator (\ref{CovariantDerivatTorsion}) can also be represented as:
\begin{eqnarray}
\hat {\tilde \nabla}'_{j} = \hat \nabla_{j} - K^{l}_{kj}\chi^k \frac\partial{\partial \chi^l},
\end{eqnarray}
where
\begin{eqnarray}
K^{l}_{jk} = \tilde \Gamma^{l}_{jk} - \Gamma^{l}_{jk},
\end{eqnarray}
is the cotorsion tensor,
\begin{eqnarray}
\Gamma^l_{jk} = \frac12g^{lp}\left(g_{pk'j}+g_{pj'k} - g_{jk'p}\right),
\end{eqnarray}
is the Levi-Civita connection (we introduced $(...)_{'j}\equiv \partial (...)/\partial x^j$), and
\begin{eqnarray}
\hat \nabla_{j} = \frac\partial{\partial x^j} - \Gamma_{jk}^l\chi^k \frac\partial{\partial \chi^l},
\end{eqnarray}
is its covariant derivative operator.

It is tempting to believe that the diffusion Laplacian of stochastic quantization is the Hodge Laplacian, $[\hat d, \hat d^\dagger]$, where $\hat d^\dagger$ is the adjoint of $\hat d$. To see if this is true, we recall that the definition of the adjoint is $\langle \phi | \hat d \psi \rangle = \langle \hat d^\dagger \phi | \psi \rangle, \forall \psi,\phi$, where the scalar product $\langle \phi | \psi \rangle = \int_X \psi \star \psi^*$, and the Hodge conjugation
\begin{eqnarray}
\star (\psi_{i_1...i_r}\chi^{i_1}...\chi^{i_r}) = \frac1{(D-r)!}\psi_{i'_1...i'_r}g^{1/2}
g^{i'_1i_1}...g^{i'_ri_r}\epsilon_{i_1...i_D}
\chi^{i_{r+1}}...\chi^{i_D},
\end{eqnarray}
with $g=\det g_{ij}$ and $\epsilon_{i_1...i_D}$ being the antisymmetric tensor. Equipped with these definitions and other useful formulas such as:
\begin{eqnarray}
\epsilon_{i_1...i_ri_{r+1}...i_D}\epsilon_{j_1...j_ri_{r+1}...i_D} = (D-r)!
\left(\begin{array}{ccc}
  \delta_{i_1j_1} & \dots & \delta_{i_1j_r} \\
  \vdots &  \ddots & \vdots \\
  \delta_{i_rj_1}&\dots&\delta_{i_rj_r}\\
\end{array}\right),
\end{eqnarray}
one can derive a few well-known relations such as $\star\star = (-1)^{(D-\hat F)\hat F}$,  $(\chi^i)^\dagger=g^{ij}\partial/\partial\chi^j$, and eventually
\begin{eqnarray}
\hat d^\dagger &=& \left(\frac\partial{\partial x^i}\chi^i\right)^\dagger = -\frac\partial{\partial \chi^i} g^{ij}\left(-\frac\partial{\partial x^j}\right)^\dagger,\label{ddagger1}
\end{eqnarray}
with
\begin{eqnarray}
\left(-\frac\partial{\partial x^j}\right)^\dagger&=& \hat \nabla_{j} + g_{kr}\Gamma^{r}_{jp} g^{pl} \frac\partial{\partial \chi^l} \chi^k.\label{ddagger2}
\end{eqnarray}
Comparing Eqs.(\ref{ddagger1}) and (\ref{ddagger2}) with Eqs.(\ref{adjoint}) and (\ref{CovariantDerivatTorsion}), we see that
\begin{eqnarray}
\hat{\bar d} \ne \hat d^\dagger.
\end{eqnarray}
This inequality, however, is not quite the same as $[\hat d, \hat {\bar d} ] \ne [\hat d, \hat d^\dagger]$. Strictly speaking, the diffusion Laplacian equals the Hodge Laplacian if
\begin{eqnarray}
[\hat d, \hat{\bar d} - \hat d^\dagger]=\left[\hat d, \frac\partial{\partial \chi^i} g^{ij}\left( \left(-\frac\partial{\partial x^j}\right)^\dagger -\hat{\tilde \nabla}'_j\right) \right]=0.\label{ConditionLaplacian}
\end{eqnarray}
To check if this holds, we introduce for brevity
\begin{eqnarray}
Y^l_{jk}  = K^{l}_{kj} - g_{kr}\Gamma^{r}_{jp}g^{pl},
\end{eqnarray}
so that
\begin{eqnarray}
\left(-\frac\partial{\partial x^j}\right)^\dagger-\hat{\tilde \nabla}'_j&=& Y^l_{jk}\chi^k\frac\partial{\partial \chi^l} + (\log \sqrt g)_{'j},
\end{eqnarray}
where we used $\Gamma^k_{jk} = (\log \sqrt g)_{'j}$. Thus,
\begin{eqnarray}
\left[\hat d, \hat{\bar d} - \hat d^\dagger\right] &=& \left[\hat d, g^{ij}Y^l_{jk}\frac\partial{\partial \chi^i} \chi^k\frac\partial{\partial \chi^l} + g^{ij}(\log \sqrt{g})_{'j}\frac\partial{\partial \chi^i}\right]\nonumber\\
&=& \Big((g^{ij}Y^l_{jk})_{'m}\Big)\chi^m\frac{\partial}{\partial \chi^i}\chi^k\frac{\partial}
{\partial\chi^l} + \Big(g^{ij}Y^l_{jk} - g^{lj}Y^i_{jk}\Big) \chi^k\frac{\partial}{\partial \chi^l}\frac{\partial}{\partial x^i}\nonumber\\
&& + \Big(g^{ij}(\log \sqrt g)_{'j} + g^{lj}Y^i_{jl}\Big)
\frac{\partial}{\partial x^i} +\Big((g^{ij}(\log\sqrt g)_{'j})_{'m}\Big)\chi^m\frac{\partial }{\partial \chi^i}.\nonumber
\end{eqnarray}
For the r.h.s. of this equation to vanish in the operator sense, each of the four coefficients in the big brackets must equal zero. This is not so in the general case as is clearly seen from the fourth coefficient for instance. Thus, we came into conclusion that in general the diffusion Laplacian is not the Hodge Laplacian. This finding is rather surprising. It suggests that supersymmetric nonlinear sigma models (see, e.g., Chapter 10.4 of Ref.\cite{MirrorSymmetry}) that all have Hodge Laplacians are not exactly the Riemannian phase-space generalization of the stochastic quantization of Langevin SDEs.

The diffusion Laplacian is the Hodge Laplacian only for some classes of models. One such class is trivial - it is the models with flat metric (additive noise), \emph{i.e.}, with coordinate-independent veilbeins and consequently vanishing connection. Another such class is the torsion-free models. To see that this is true one can use an alternative representation of the diffusion Laplacian
\begin{eqnarray}
[\hat d, \hat {\bar d}] = - \left[\hat d, \frac\partial{\partial\chi^i_a}e^i_a(x)\right]\left[\hat d, \frac\partial{\partial\chi^j}e^j_a(x)\right] = -(\hat{\tilde \nabla}'_i+\tilde\Gamma^l_{li}) g^{ij}\hat{\tilde\nabla}'_i.\label{AlternatLapl}
\end{eqnarray}
Now, if torsion is zero, $\tilde\Gamma^i_{jk}=\Gamma^i_{kj}$, $\hat{\tilde\nabla}'_i=\hat{\nabla}_i$, $-(\hat{\tilde \nabla}'_i+\tilde\Gamma^l_{li})=-(\hat{ \nabla}_i+\Gamma^l_{il})=\hat{\nabla}_i^\dagger$ and the diffusion Laplacian is the Bochner Laplacian, $\hat \nabla^\dagger_i g^{ij}\hat \nabla_j$. It remains now to recall that according to the Weitzenb\"ock formula, the Hodge Laplacian equals the Bochner Laplacian when the Riemann curvature tensor (of the Levi-Chivita connection) is zero, which is indeed the case for the vanishing torsion of the Wietzenb\"ock connection (see, \emph{e.g.}, Eqs. (49) and (50) of Ref.\cite{TeleparallelGravity}).

Finalizing the discussion in this section, we would like to address here the following issue. In the literature on SDEs, the Weyl quantization technique is known as Stratonovich approach, which is often opposed to the Ito approach. The later uses $x_0$ instead of $(x_1+x_0)/2$ in Eq.(\ref{Eq2Noname}) and corresponds to the unphysical convention of evaluating the flow vector field in the r.h.s of the discrete-time approximation of the SDE at the very beginning of the interval of the infinitesimal time-evolution, $(0,\Delta t)$. On the level of the operator ordering, Ito approach is equivalent to the rule that all the momentum operators are on the "left" of (or act after) all the position operators. This operator ordering rule leads to a different Fokker-Planck operator:
\begin{eqnarray}
\hat H_{Ito} = [\hat d, \hat j_{Ito}], {  }\hat j_{Ito} = \hat j - \Theta \tilde \Gamma^i_{kj}g^{jk}\frac\partial{\partial \chi^i}.\label{ItoFP}
\end{eqnarray}
At the same time, it is understood that there can be only one correct evolution operator and the question that we need to address now is which of the two evolution operators is correct. This question is known in the literature as the Ito-Stratonovich dilemma. \cite{VonKampen}

Several reasons have been found pointing onto why the Weyl-Stratonovich approach to SDEs is more reasonable (see, e.g., Refs. \cite{Moon} and Refs. therein). Moreover, it was demonstrated that the results obtained within the Weyl-Stratonovich view on SDEs are closer to experimental data. \cite{PhysicalProofOfStratonovich} Furthermore, if Ito operator ordering rule is applied to an ordinary quantization of a non-relativistic particle of mass, $m$, in a magnetic field defined by a vector potential, $A$, it leads to a non-Hermitian and non-gauge-invariant kinetic part of the Hamiltonian: $ (\hat p^2 - 2\hat p A + A^2)/(2m)\ne (\hat p^2 - 2\hat p A + A^2)^\dagger/(2m), { } (\hat p -A)^2/(2m)$, where $\hat p = -i \hbar \partial/\partial x$ is the momentum operator and $\hbar$ is the Planck constant. This is yet another reason why Ito approach is unphysical.

Getting back to our discussion, the pathintegral technique we used so far does not resolve the Ito-Stratonovich dilemma. Indeed, the Ito evolution operator (\ref{ItoFP}) is still $\hat d$-exact and consequently there is still some room to believe that Ito approach is acceptable. In order to resolve this ambiguity one can proceed as follows.

Let us return to the infinitesimal temporal evolution introduced in the beginning of this section. The infinitesimal pullback in Eq.(\ref{Eq1Noname}) can be alternatively given as the exponentiation of the Lie derivative:
\begin{eqnarray}
\psi(x\chi(t+\Delta t)) &=& \langle e^{-\Delta t \hat{\mathcal{L}}_{F(\xi)}} \rangle_{Ns} \psi(x\chi t)\nonumber\\&=&\left\langle 1 -\Delta t \hat{\mathcal{L}}_{F(\xi)} + \frac12 \Delta t^2 \hat{\mathcal{L}}_{F(\xi)} \hat{\mathcal{L}}_{F(\xi)} + ... \right\rangle_{Ns} \psi(x\chi t), \label{IntinitesimalLie}\\
\hat{\mathcal{L}}_{F(\xi)} &=& \hat{\mathcal{L}}_{F} + (2T)^{1/2} \xi^a \hat{\mathcal{L}}_{e_a} = \left[\hat d, \frac\partial{\partial\chi^i}\left(F^i + (2\Theta)^{1/2} e^i_a \xi^a\right) \right].
\end{eqnarray}
Unlike in Eq.(\ref{Eq1Noname}), in the operator version of the infinitesimal evolution one need not to separate the variables into those living on the first and the second time-slices ($(x_1\chi_1),(x_0\chi_0)\to (x\chi),\xi_0\to \xi$).

Using $\langle\xi^a\xi^b\rangle_{Ns} = \delta^{ab}/\Delta t$ and $\langle\xi^{a_1}...\xi^{a_{2k}}\rangle_{Ns}\propto (\Delta t)^{-k}$, and keeping in Eq.(\ref{IntinitesimalLie}) only terms linear in $\Delta t$, one arrives at
\begin{eqnarray}
\partial_t \psi(x\chi t) = \lim_{\Delta t\to0}\frac{\psi(x\chi(t+\Delta t))-\psi(x\chi t)}{\Delta t} = - \hat H\psi(x\chi t),\label{FPEqNew1}
\end{eqnarray}
where the evolution operator is
\begin{eqnarray}
\hat H  = \hat{\mathcal{L}}_{F} - \Theta \hat{\mathcal{L}}_{e_a}\hat{\mathcal{L}}_{e_a} = \left[\hat d, \frac\partial{\partial\chi^i}F^i(x)\right] - \Theta \left[\hat d, \frac\partial{\partial\chi^i}e^i_a(x)\right]\left[\hat d, \frac\partial{\partial\chi^j}e^j_a(x)\right].\label{FPEqNew2}
\end{eqnarray}
Eq.(\ref{FPEqNew1}) is the operator version of Eq.(\ref{WeylIntegral}), whereas Eq.(\ref{FPEqNew2}) is the Weyl-Stratonovich Fokker-Planck operator as is seen from Eqs.(\ref{HamiltAdditional}), (\ref{jadditional}), and (\ref{AlternatLapl}). For a total probability distribution, $P(x\chi t) = P(xt)\chi^1...\chi^D \in\Omega^D(X)$, one has $\hat d P(x\chi t)=0$, $[\hat d, (\partial/\partial\chi^j)e^j_a(x)]P(x\chi t) = (\partial/\partial x^j)e^j_a(x) P(x\chi t)$, and the Fokker-Planck equation takes the form of the conventional Fokker-Planck equation in the Stratonovich approach:
\begin{eqnarray}
\partial_t P(xt) = \left( - \frac{\partial}{\partial x^i}F^i  + \Theta \frac{\partial}{\partial x^i}e^i_a \frac{\partial}{\partial x^j}e^j_a \right)P(xt).
\end{eqnarray}

No approximations were made in the derivation of Eqs.(\ref{FPEqNew1}) and (\ref{FPEqNew2}). This unambiguously resolves the Ito-Stratonovich dilemma in favor of the Weyl-Stratonovich approach. What resolves the issue here is the correct operator ordering provided by the understanding that the infinitesimal evolution/pullback is the Lie derivative (see Eq.(\ref{IntinitesimalLie})).




\end{document}